\pdfoutput=1
\documentclass[12pt,a4paper]{article}

\usepackage{ifthen} 

\usepackage{ifthen} 
\usepackage{rotating}
\usepackage{dashrule}
\usepackage{array} 
\usepackage{dcolumn}
\usepackage{comment}
\usepackage{xcolor}
\usepackage{graphpap} 
\usepackage{rotating} 
\usepackage{tikz}
\usepackage{xcolor}
\usepackage{longtable} 
\usepackage{lscape}
\usepackage{multirow} 
\usepackage{mathrsfs}
\usepackage{mathtools} 
\usetikzlibrary{patterns}
\usepackage{nicefrac} 
\usepackage{transparent}
\usepackage{afterpage} 
\usepackage{comment} 
\usepackage[normalem]{ulem}
\usepackage{cancel}

\newboolean{pdflatex}
\setboolean{pdflatex}{true} 

\newboolean{articletitles}
\setboolean{articletitles}{true} 

\newboolean{uprightparticles}
\setboolean{uprightparticles}{true} 

\def\paperauthors{LHCb collaboration} 
\def\paperasciititle{Template for writing LHCb papers} 
\def\papertitle{Observation of 
the~$\decay{\Bp}{\jpsi\Peta^{\prime}\Kp}$~decay} 
\def\paperkeywords{{High Energy Physics}, {LHCb}} 
\def\papercopyright{\the\year\ CERN for the benefit of the LHCb collaboration} 
\def\paperlicence{CC BY 4.0 licence}
\def\paperlicenceurl{https://creativecommons.org/licenses/by/4.0/}

\DeclareMathOperator*{\bigplus}{\scalerel*{+}{\sum}}
\usepackage{scalerel}

\usepackage[top=1in, bottom=1.25in, left=1in, right=1in]{geometry}

\usepackage{microtype}
\usepackage{lineno}  
\usepackage{xspace} 
\usepackage{caption} 

\usepackage{graphicx}  
\usepackage{color}
\usepackage{colortbl}
\graphicspath{{./figs/}} 

\usepackage{amsmath} 
\usepackage{amssymb}
\usepackage{amsfonts}
\usepackage{upgreek} 
\usepackage{multirow}
\newcommand*\patchAmsMathEnvironmentForLineno[1]{%
\expandafter\let\csname old#1\expandafter\endcsname\csname #1\endcsname
\expandafter\let\csname oldend#1\expandafter\endcsname\csname
end#1\endcsname
 \renewenvironment{#1}%
   {\linenomath\csname old#1\endcsname}%
   {\csname oldend#1\endcsname\endlinenomath}%
}
\newcommand*\patchBothAmsMathEnvironmentsForLineno[1]{%
  \patchAmsMathEnvironmentForLineno{#1}%
  \patchAmsMathEnvironmentForLineno{#1*}%
}
\AtBeginDocument{%
\patchBothAmsMathEnvironmentsForLineno{equation}%
\patchBothAmsMathEnvironmentsForLineno{align}%
\patchBothAmsMathEnvironmentsForLineno{flalign}%
\patchBothAmsMathEnvironmentsForLineno{alignat}%
\patchBothAmsMathEnvironmentsForLineno{gather}%
\patchBothAmsMathEnvironmentsForLineno{multline}%
\patchBothAmsMathEnvironmentsForLineno{eqnarray}%
}

\usepackage{hyperxmp}

\usepackage[pdftex,
            pdfauthor={\paperauthors},
            pdftitle={\paperasciititle},
            pdfkeywords={\paperkeywords},
            pdfcopyright={Copyright (C) \papercopyright},
            pdflicenseurl={\paperlicenceurl}]{hyperref}

\usepackage[colorinlistoftodos,textsize=scriptsize]{todonotes}

\usepackage[bottom,flushmargin,hang,multiple]{footmisc}

\usepackage[all]{hypcap} 

\usepackage{xspace} 
\usepackage{upgreek}

\def\lhcb   {\mbox{LHCb}\xspace}

\def\belle  {\mbox{Belle}\xspace}

\def\MagUp {\mbox{\em Mag\kern -0.05em Up}\xspace}

\ifthenelse{\boolean{uprightparticles}}%
{
 
 \def\Pgamma      {\ensuremath{\upgamma}\xspace}

 \def\Pvarepsilon {\ensuremath{\upvarepsilon}\xspace}                 
                  
 \def\Peta        {\ensuremath{\upeta}\xspace}

 \def\Pmu         {\ensuremath{\upmu}\xspace}

 \def\Ppi         {\ensuremath{\uppi}\xspace}                 
                  
 \def\Prho        {\ensuremath{\uprho}\xspace}

 \def\Pphi        {\ensuremath{\upphi}\xspace}                 
                  
 \def\Pchi        {\ensuremath{\upchi}\xspace}                 
 \def\Ppsi        {\ensuremath{\uppsi}\xspace}                 
 \def\Pomega      {\ensuremath{\upomega}\xspace}                 

 \def\PDelta      {\ensuremath{\Delta}\xspace}                 
 \def\PXi         {\ensuremath{\Xi}\xspace}                 
 \def\PLambda     {\ensuremath{\Lambda}\xspace}                 
 \def\PSigma      {\ensuremath{\Sigma}\xspace}                 
 \def\POmega      {\ensuremath{\Omega}\xspace}                 
 \def\PUpsilon    {\ensuremath{\Upsilon}\xspace}
 \let\oldPi\Pi
 \def\PPi         {\ensuremath{\oldPi}\xspace}

 \def\PB      {\ensuremath{\mathrm{B}}\xspace}                 
                  
 \def\PD      {\ensuremath{\mathrm{D}}\xspace}

 \def\PJ      {\ensuremath{\mathrm{J}}\xspace}                 
 \def\PK      {\ensuremath{\mathrm{K}}\xspace}

 \def\PS      {\ensuremath{\mathrm{S}}\xspace}

 \def\Pb      {\ensuremath{\mathrm{b}}\xspace}                 
 \def\Pc      {\ensuremath{\mathrm{c}}\xspace}                 
                  
 \def\Pe      {\ensuremath{\mathrm{e}}\xspace}

 \def\Pi      {\ensuremath{\mathrm{i}}\xspace}

 \def\Pp      {\ensuremath{\mathrm{p}}\xspace}

 \def\Ps      {\ensuremath{\mathrm{s}}\xspace}

 \def\thebaroffset{0.0em}
}
{
 
 \def\Pgamma      {\ensuremath{\gamma}\xspace}

 \def\Pvarepsilon {\ensuremath{\varepsilon}\xspace}                 
                  
 \def\Peta        {\ensuremath{\eta}\xspace}

 \def\Pmu         {\ensuremath{\mu}\xspace}

 \def\Ppi         {\ensuremath{\pi}\xspace}                 
                  
 \def\Prho        {\ensuremath{\rho}\xspace}

 \def\Pphi        {\ensuremath{\phi}\xspace}                 
                  
 \def\Pchi        {\ensuremath{\chi}\xspace}                 
 \def\Ppsi        {\ensuremath{\psi}\xspace}                 
 \def\Pomega      {\ensuremath{\omega}\xspace}                 
 \mathchardef\PDelta="7101
 \mathchardef\PXi="7104
 \mathchardef\PLambda="7103
 \mathchardef\PSigma="7106
 \mathchardef\POmega="710A
 \mathchardef\PUpsilon="7107
 \mathchardef\PPi="7105
                  
 \def\PB      {\ensuremath{B}\xspace}                 
                  
 \def\PD      {\ensuremath{D}\xspace}

 \def\PJ      {\ensuremath{J}\xspace}                 
 \def\PK      {\ensuremath{K}\xspace}

 \def\PS      {\ensuremath{S}\xspace}

 \def\Pb      {\ensuremath{b}\xspace}                 
 \def\Pc      {\ensuremath{c}\xspace}                 
                  
 \def\Pe      {\ensuremath{e}\xspace}

 \def\Pi      {\ensuremath{i}\xspace}

 \def\Pp      {\ensuremath{p}\xspace}

 \def\Ps      {\ensuremath{s}\xspace}

 \def\thebaroffset{0.18em}
}
\newcommand{\offsetoverline}[2][\thebaroffset]{\kern #1\overline{\kern -#1 #2}}%

\makeatletter
\ifcase \@ptsize \relax
  \newcommand{\miniscule}{\@setfontsize\miniscule{4}{5}}
\or
  \newcommand{\miniscule}{\@setfontsize\miniscule{5}{6}}
\or
  \newcommand{\miniscule}{\@setfontsize\miniscule{5}{6}}
\fi
\makeatother

\DeclareRobustCommand{\optbar}[1]{\shortstack{{\miniscule (\rule[.5ex]{1.25em}{.18mm})}
  \\ [-.7ex] $#1$}}

\def\epem       {{\ensuremath{\Pe^+\Pe^-}}\xspace}

\def\mumu       {{\ensuremath{\Pmu^+\Pmu^-}}\xspace}

\def\g      {{\ensuremath{\Pgamma}}\xspace}

\def\squark    {{\ensuremath{\Ps}}\xspace}

\def\cquark    {{\ensuremath{\Pc}}\xspace}

\def\bquark    {{\ensuremath{\Pb}}\xspace}

\def\pion   {{\ensuremath{\Ppi}}\xspace}
\def\piz    {{\ensuremath{\pion^0}}\xspace}
\def\pip    {{\ensuremath{\pion^+}}\xspace}
\def\pim    {{\ensuremath{\pion^-}}\xspace}

\def\rhomeson {{\ensuremath{\Prho}}\xspace}
\def\rhoz     {{\ensuremath{\rhomeson^0}}\xspace}

\def\kaon    {{\ensuremath{\PK}}\xspace}

\def\KorKbar {\kern \thebaroffset\optbar{\kern -\thebaroffset \PK}{}\xspace}

\def\Kp      {{\ensuremath{\kaon^+}}\xspace}
\def\Km      {{\ensuremath{\kaon^-}}\xspace}

\def\KS      {{\ensuremath{\kaon^0_{\mathrm{S}}}}\xspace}

\def\Kstarp  {{\ensuremath{\kaon^{*+}}}\xspace}

\newcommand{\etapr}{\ensuremath{\Peta^{\prime}}\xspace}

\def\D       {{\ensuremath{\PD}}\xspace}

\def\DorDbar {\kern \thebaroffset\optbar{\kern -\thebaroffset \PD}\xspace}
\def\Dz      {{\ensuremath{\D^0}}\xspace}

\def\Dp      {{\ensuremath{\D^+}}\xspace}
\def\Dm      {{\ensuremath{\D^-}}\xspace}

\def\DpDm    {\ensuremath{\Dp {\kern -0.16em \Dm}}\xspace}

\def\Dstarp  {{\ensuremath{\D^{*+}}}\xspace}

\def\B       {{\ensuremath{\PB}}\xspace}

\def\BorBbar {\kern \thebaroffset\optbar{\kern -\thebaroffset \PB}\xspace}

\def\Bd      {{\ensuremath{\B^0}}\xspace}

\def\BdorBdbar {\kern \thebaroffset\optbar{\kern -\thebaroffset \Bd}\xspace}
\def\Bu      {{\ensuremath{\B^+}}\xspace}

\def\Bp      {{\ensuremath{\Bu}}\xspace}

\def\Bs      {{\ensuremath{\B^0_\squark}}\xspace}

\def\BsorBsbar {\kern \thebaroffset\optbar{\kern -\thebaroffset \Bs}\xspace}

\def\jpsi     {{\ensuremath{{\PJ\mskip -3mu/\mskip -2mu\Ppsi}}}\xspace}
\def\psitwos  {{\ensuremath{\Ppsi{(2S)}}}\xspace}

\def\chicone  {{\ensuremath{\Pchi_{\cquark 1}}}\xspace}

\def\Y#1S{\ensuremath{\PUpsilon{(#1S)}}\xspace}

\def\proton      {{\ensuremath{\Pp}}\xspace}

\def\Lz          {{\ensuremath{\PLambda}}\xspace}

\def\LorLbar     {\kern \thebaroffset\optbar{\kern -\thebaroffset \PLambda}\xspace}

\def\BF         {{\ensuremath{\mathcal{B}}}\xspace}
\def\BR         {\BF}

\newcommand{\decay}[2]{\ensuremath{#1\!\to #2}\xspace} 

\def\to                 {\ensuremath{\rightarrow}\xspace}

\def\AT#1     {\ensuremath{A_{\mathrm{T}}^{#1}}\xspace}           

\def\C#1      {\ensuremath{\mathcal{C}_{#1}}\xspace}                       
\def\Cp#1     {\ensuremath{\mathcal{C}_{#1}^{'}}\xspace}                    
\def\Ceff#1   {\ensuremath{\mathcal{C}_{#1}^{\mathrm{(eff)}}}\xspace}        
\def\Cpeff#1  {\ensuremath{\mathcal{C}_{#1}^{'\mathrm{(eff)}}}\xspace}       
\def\Ope#1    {\ensuremath{\mathcal{O}_{#1}}\xspace}                       
\def\Opep#1   {\ensuremath{\mathcal{O}_{#1}^{'}}\xspace}                    

\newcommand{\nospaceunit}[1]{\ensuremath{\text{#1}}}       
\newcommand{\aunit}[1]{\ensuremath{\text{\,#1}}}       

\newcommand{\tev}{\aunit{Te\kern -0.1em V}\xspace}
\newcommand{\gev}{\aunit{Ge\kern -0.1em V}\xspace}
\newcommand{\mev}{\aunit{Me\kern -0.1em V}\xspace}
\newcommand{\kev}{\aunit{ke\kern -0.1em V}\xspace}
\newcommand{\ev}{\aunit{e\kern -0.1em V}\xspace}
 
\newcommand{\mevc}{\ensuremath{\aunit{Me\kern -0.1em V\!/}c}\xspace}
\newcommand{\gevc}{\ensuremath{\aunit{Ge\kern -0.1em V\!/}c}\xspace}
\newcommand{\mevcc}{\ensuremath{\aunit{Me\kern -0.1em V\!/}c^2}\xspace}
\newcommand{\gevcc}{\ensuremath{\aunit{Ge\kern -0.1em V\!/}c^2}\xspace}

\def\mum  {\ensuremath{\,\upmu\nospaceunit{m}}\xspace}

\def\fb   {\ensuremath{\aunit{fb}}\xspace}
\def\invfb   {\ensuremath{\fb^{-1}}\xspace}

\newcommand{\chisq}{\ensuremath{\chi^2}\xspace}

\def\gsim{{~\raise.15em\hbox{$>$}\kern-.85em
          \lower.35em\hbox{$\sim$}~}\xspace}
\def\lsim{{~\raise.15em\hbox{$<$}\kern-.85em
          \lower.35em\hbox{$\sim$}~}\xspace}

\def\sPlot{\mbox{\em sPlot}\xspace}

\def\pt         {\ensuremath{p_{\mathrm{T}}}\xspace}

\def\ptot       {\ensuremath{p}\xspace}

\def\evtgen     {\mbox{\textsc{EvtGen}}\xspace}

\def\geant      {\mbox{\textsc{Geant4}}\xspace}

\def\photos     {\mbox{\textsc{Photos}}\xspace}

\def\pythia     {\mbox{\textsc{Pythia}}\xspace}

\def\tell1  {TELL1\xspace}
\def\ukl1   {UKL1\xspace}

\newcommand{\lhcborcid}[1]{\href{https://orcid.org/#1}{\hspace*{0.1em}\raisebox{-0.45ex}{\includegraphics[width=1em]{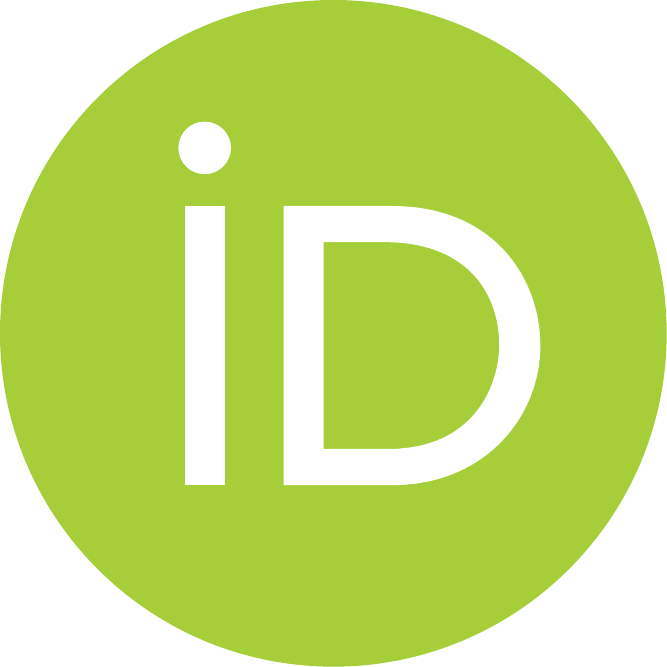}}}}

\usepackage{cite} 
\usepackage{mciteplus}

\usepackage{longtable} 
\usepackage{comment}

\makeatletter
\g@addto@macro\bfseries{\boldmath}
\makeatother

\begin{document}

\def\betapk{\mbox{\decay{\Bp}{\jpsi\etapr\Kp}}}
\def\betap{\mbox{\decay{\Bs}{\jpsi\etapr}}}
\def\etaprhog{\mbox{\decay{\etapr}{\Prho^0\g}}}
\def\etapetapipi{\mbox{\decay{\etapr}{\Peta\pip\pim}}}

\def\bpipik{\mbox{\decay{\Bp}{\jpsi\pip\pim\Kp}}}
\def\bpsik{\mbox{\decay{\Bp}{\psitwos\Kp}}}
\def\psipipi{\jpsi\pip\pim}

\def\psietap{\ensuremath{\jpsi\etapr}\xspace}
\def\psik{\ensuremath{\jpsi\Kp}\xspace}
\def\etapk{\ensuremath{\etapr\Kp}\xspace}
\def\MLP{\ensuremath{\mathrm{MLP}}\xspace}

\def\efftot{\ensuremath{\Pvarepsilon^{\mathrm{tot}}}\xspace}
\def\effgen{\ensuremath{\Pvarepsilon^{\mathrm{gen}}}\xspace}
\def\effrec{\ensuremath{\Pvarepsilon^{\mathrm{reco\&presel}}}\xspace}
\def\effmlp{\ensuremath{\Pvarepsilon^{\MLP}}\xspace}
\def\efftrig{\ensuremath{\Pvarepsilon^{\mathrm{trig}}}\xspace}

\renewcommand{\thefootnote}{\fnsymbol{footnote}}
\setcounter{footnote}{1}


\begin{titlepage}
\pagenumbering{roman}

\vspace*{-1.5cm}
\centerline{\large EUROPEAN ORGANIZATION FOR NUCLEAR RESEARCH (CERN)}
\vspace*{1.5cm}
\noindent
\begin{tabular*}{\linewidth}{lc@{\extracolsep{\fill}}r@{\extracolsep{0pt}}}
\ifthenelse{\boolean{pdflatex}}
{\vspace*{-1.5cm}\mbox{\!\!\!\includegraphics[width=.14\textwidth]{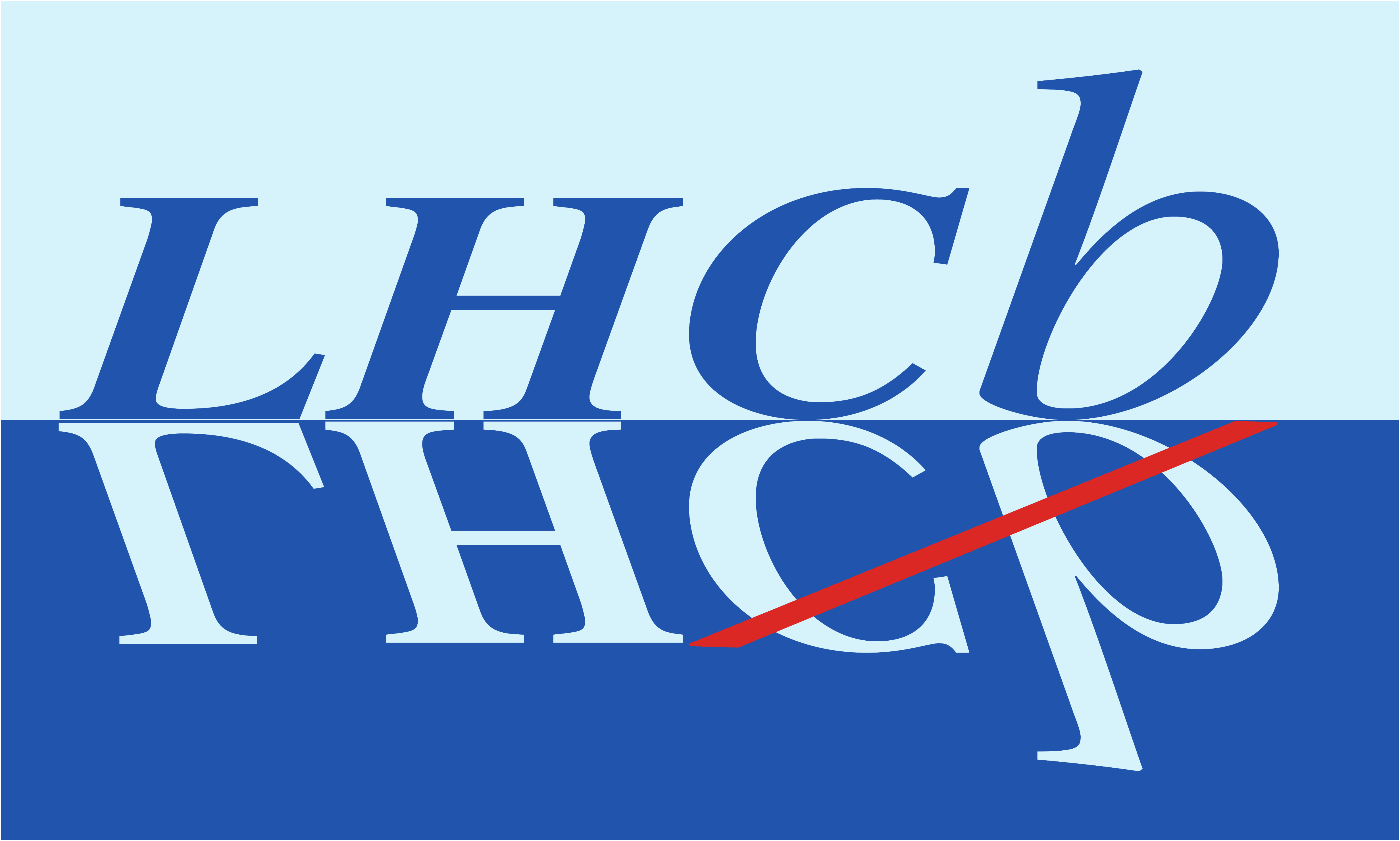}} & &}%
{\vspace*{-1.2cm}\mbox{\!\!\!\includegraphics[width=.12\textwidth]{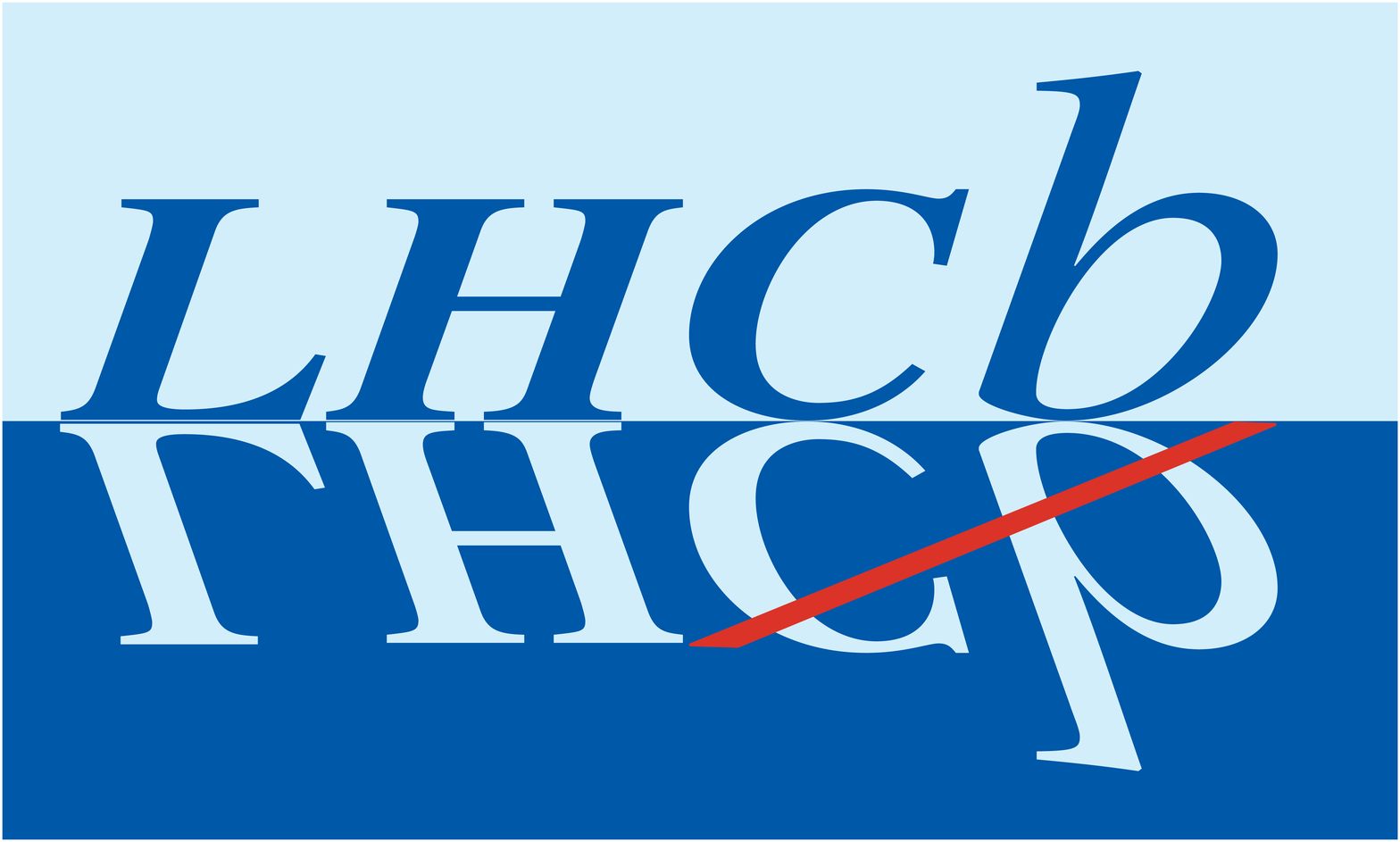}} & &}%
\\
 & & CERN-EP-2023-022 \\  
 & & LHCb-PAPER-2022-054 \\  
 & & March 13,  2023 \\ 
 & & \\
\end{tabular*}

\vspace*{3.0cm}

{\normalfont\bfseries\boldmath\huge
\begin{center}
  \papertitle 
\end{center}
}

\vspace*{1.2cm}

\begin{center}
\paperauthors\footnote{Authors are listed at the end of this paper.}
\end{center}

\vspace{\fill}

\begin{abstract}
  \noindent
  The \betapk decay is observed for the first time 
  using 
  proton\nobreakdash-proton collision data
  collected by the~LHCb experiment  
  at centre-of-mass energies of 
  7, 8, and 13\tev,
  corresponding 
  to a~total integrated luminosity of $9\invfb$.
  The~branching fraction of this decay is measured relative 
  to the known branching fraction
  of the~\bpsik~decay
  and found to be 
  $$
  \dfrac{\BF(\betapk)}{\BF(\bpsik)} = 
  \left(4.91\pm 0.47\pm0.29\pm0.07\right)\times10^{-2},
  $$
  where the~first uncertainty is statistical, 
  the~second is systematic and 
  the~third is related to
  external branching fractions. 
  A~first look at the~\psietap 
  mass distribution 
  is performed 
  and no signal of intermediate 
  resonances is observed.
  \end{abstract}

\vspace*{2.0cm}

\begin{center}
  Published  in  \href{https://doi.org/10.1007/JHEP08(2023)174}{JHEP 08 (2023) 174}
\end{center}

\vspace{\fill}

{\footnotesize 
\centerline{\copyright~\papercopyright. \href{\paperlicenceurl}{\paperlicence}.}}
\vspace*{2mm}

\end{titlepage}


\newpage
\setcounter{page}{2}
\mbox{~}


\renewcommand{\thefootnote}{\arabic{footnote}}
\setcounter{footnote}{0}

\cleardoublepage


\pagestyle{plain} 
\setcounter{page}{1}
\pagenumbering{arabic}


\section{Introduction}
\label{sec:intro}

In~the~last twenty 
years 
a~plethora of new hadron states
have been discovered 
in decays of beauty hadrons 
to
    charmonium,
including 
the~enigmatic $\chicone(3872)$~state~\cite{Choi:2003ue}, 
numerous pentaquark 
states in $\jpsi\proton$~\cite{LHCb-PAPER-2015-029,
LHCb-PAPER-2016-009,
LHCb-PAPER-2019-014,
LHCb-PAPER-2016-015,
LHCb-PAPER-2021-018}
and $\jpsi\Lz$~\cite{LHCb-PAPER-2020-039}~systems
as well as tetraquarks  
in 
the~$\psitwos\pip$~\cite{Choi:2007wga,
Mizuk:2009da,
Chilikin:2013tch,
LHCb-PAPER-2014-014,
LHCb-PAPER-2015-038},
$\jpsi\Pphi$~\cite{LHCb-PAPER-2016-018,
LHCb-PAPER-2016-019,
LHCb-PAPER-2020-035,
LHCb-PAPER-2020-044,
LHCb-PAPER-2022-040},
$\Peta_{\cquark}(1\PS)\pim$~\cite{LHCb-PAPER-2018-034},
$\jpsi\pip$~\cite{LHCb-PAPER-2018-043},
$\jpsi\Kp$~\cite{LHCb-PAPER-2020-044}
and 
$\jpsi\KS$~\cite{LHCb-PAPER-2022-040}~systems. 
Transitions among
    charmonium or charmonium\nobreakdash-like
states have been studied in 
beauty-hadron decays,  
including transitions with emission of 
one~photon~\cite{LHCb-PAPER-2013-024,
LHCb-PAPER-2014-008,
LHCb-PAPER-2017-011,
LHCb-PAPER-2021-003}, 
two pions~\cite{LHCb-PAPER-2013-001,
LHCb-PAPER-2015-015,
LHCb-PAPER-2020-009, 
LHCb-PAPER-2021-045},
$\Pphi$~\cite{LHCb-PAPER-2016-018,
LHCb-PAPER-2016-019,
LHCb-PAPER-2020-035,
LHCb-PAPER-2020-044,
LHCb-PAPER-2022-040},
$\Pomega$~\cite{BaBar:2010wfc,
LHCb-PAPER-2021-045} 
and $\Peta$~\cite{LHCb-PAPER-2021-047}~mesons.
The~transitions with emission of 
an~\Peta~meson 
have 
also 
been
studied in 
\mbox{$\decay{\epem}
{\jpsi\Peta}$}~processes~\cite{Wang:2012bgc,
BESIII:2020bgb}. 
In~general, studies of various hadronic transitions
in
the~charmonium and charmonium\nobreakdash-like
sectors 
can shed light
    onto
the~internal 
structure of these particles, 
which is largely unknown for newly 
discovered hadronic states~\cite{Brambilla:2019esw,
Ali:2019roi,
Maiani:2022psl,
Brambilla:2022ura}.

Transitions with an~emission of 
an~$\Peta^{\prime}$~meson in
    the~charmonium and charmonium\nobreakdash-like
systems have not yet 
been observed~\cite{Brambilla:2010cs,
Brambilla:2019esw,
PDG2022}. 
Since the~\etapr~meson may have 
a~glueball contribution~\cite{Rosner:1982ey,
Bramon:1997va,
Bramon:1997mf,
Novikov:1979uy,
Novikov:1979ux,
Novikov:1979va,
Kataev:1981aw,
Kataev:1981gr,
Thomas:2007uy,
Escribano:2007cd,
Escribano:2008rq,
Ambrosino:2009sc,
Yelton:2009aa,
PhysRevD00,
DiDonato:2011kr,
LHCb-PAPER-2014-056,
Andreichikov:2019ehn,
Andreichikov:2019uoa}, 
processes involving this particle are 
of particular interest~\cite{Close:1997wp,Close:2005iz}.
The~\betapk~decay\footnote{Inclusion of charge-conjugate 
states is implied throughout the~paper, 
unless otherwise 
stated.}
is a~good candidate 
to explore the~$\jpsi\etapr$~system in detail,
offering 
the~opportunity
to search 
for possible intermediate resonances. 
The~decay itself 
has never been observed
and 
an~upper limit on its branching fraction of 
\begin{equation*}
\BR(\betapk) < 8.8\times10^{-5}\ (90\,\%~\mathrm{CL})\,,
\end{equation*}
 was set 
by the~\belle collaboration~\cite{Belle:2006aam}.

This paper reports the~observation of 
the~\betapk~decay 
using 
  proton\nobreakdash-proton\,(\proton\proton) collision data
  collected by the~\lhcb experiment  
  at centre-of-mass energies of 
  7, 8, and 13\tev,
  corresponding 
  to a~total integrated luminosity of $9\invfb$.
The~measurement of its branching fraction  
normalised to 
the~well\nobreakdash-known 
branching fraction of the~$\bpsik$~decay~\cite{PDG2022}, 
\begin{equation}
\label{eq:ratio}
\mathcal{R}
\equiv
\dfrac{\BF(\betapk)}{\BF(\bpsik)} \,,
\end{equation}
is performed using
the~\etaprhog~decay.
The~observation of the~signal 
is confirmed using 
the~\etapetapipi~decay mode, 
which is also used 
as a~cross\nobreakdash-check.

\section{Detector and simulation}
\label{sec:detector}

The \lhcb detector~\cite{LHCb-DP-2008-001,LHCb-DP-2014-002} 
is a~single\nobreakdash-arm forward
spectrometer covering 
the~\mbox{pseudorapidity} range 
\mbox{$2<\eta <5$},
designed for the study of particles containing \bquark or \cquark
quarks. The detector includes 
a~high\nobreakdash-precision tracking system
consisting of 
a~silicon\nobreakdash-strip vertex detector 
surrounding 
the~$\proton\proton$
interaction region~\cite{LHCb-DP-2014-001}, 
a~large\nobreakdash-area 
silicon\nobreakdash-strip detector located
upstream of a~dipole magnet with 
a~bending power of about~$4{\mathrm{\,Tm}}$, 
and three stations of 
silicon\nobreakdash-strip detectors 
and straw drift tubes~\cite{LHCb-DP-2013-003,
LHCb-DP-2017-001}
placed downstream of the~magnet.
The~tracking system provides a~measurement of 
the~momentum, \ptot, of charged particles with
a~relative uncertainty that varies 
from~0.5\% at low momentum 
to~1.0\% at 200\gevc.
The~minimum distance of a~track 
to a~primary $\proton\proton$~collision 
vertex\,(PV), the~impact parameter  
is measured with a~resolution of~$(15+29/\pt)\mum$,
where \pt is the~component of 
the~momentum transverse to the~beam, in\,\gevc.
Different types of charged hadrons 
are distinguished using information
from two ring\nobreakdash-imaging 
Cherenkov detectors~\cite{LHCb-DP-2012-003}. 
Photons, electrons and hadrons are 
identified by a~calorimeter system consisting of
scintillating-pad and preshower detectors, 
an electromagnetic
and a~hadronic calorimeter. 
Muons are identified by 
a~system composed of alternating layers 
of iron and multiwire
proportional chambers~\cite{LHCb-DP-2012-002}.

The~online event selection 
is performed by
a~trigger~\cite{LHCb-DP-2012-004,
LHCb-DP-2019-001}, 
which consists of a~hardware stage, based on information 
from the~calorimeter and muon systems, 
followed by a~software stage, 
which applies a~full event reconstruction.
At~the~hardware trigger stage, events are required to have 
a~muon track with high transverse momentum or dimuon candidates 
in which the~product of 
the~\pt of the~muons 
has a~high value. 
In~the~software trigger, two oppositely charged muons are required 
to form a~good\nobreakdash-quality vertex that is significantly 
displaced from every~PV, with a~dimuon mass exceeding $2.7\gevcc$.

Simulated events are used to describe signal shapes 
and to compute the~efficiencies needed to determine 
the~branching fraction ratio.
In~the~simulation, $\proton\proton$~collisions are 
generated using \pythia~\cite{Sjostrand:2007gs} with 
a~specific \lhcb configuration~\cite{LHCb-PROC-2010-056}.
Decays of unstable particles are 
described by \evtgen~\cite{Lange:2001uf}, 
in which final\nobreakdash-state radiation is generated 
using \photos~\cite{davidson2015photos}.
The~interaction of the~generated particles with the~detector, 
and its response, are implemented using 
the~\geant toolkit~\cite{Allison:2006ve, *Agostinelli:2002hh} 
as described in Ref.~\cite{LHCb-PROC-2011-006}.
The~\pt~and rapidity\,($y$) 
spectra of the~$\Bu$~mesons in simulation 
are corrected
to match 
distributions in data.
The~correction factors are calculated 
by comparing the~observed
$\pt$ and $y$~spectra
for
a~high\nobreakdash-purity 
data sample of 
reconstructed \decay{\Bu}{\jpsi\Kp}~decays
with the~corresponding simulated samples. 
In~the~simulation, 
the~\mbox{$\decay{\Bu}{\jpsi\etapr\Kp}$}~decays 
are generated as phase\nobreakdash-space decays and corrected 
using  a~gradient boosted decision 
tree reweighting 
algorithm~\cite{Rogozhnikov:2016bdp}
to reproduce  the~$\jpsi\etapr$ and $\etapr\Kp$~mass
spectra 
observed in data.   
To~describe accurately the~variables used 
for kaon identification,  
the~corresponding  quantities in simulation 
are resampled according to values obtained from
calibration data samples of 
\mbox{$\decay{\Dstarp}
{\left( \decay{\Dz}{\Km\pip}\right)\pip}$}~decays~\cite{LHCb-DP-2018-001}. 
The~procedure accounts for correlations between 
the~variables associated
    with
a~particular track, 
as well as the~dependence of 
the~kaon identification response
on
the~track's $\pt$ and  $\eta$~and  the~multiplicity 
of tracks in the~event. 
To~account for imperfections in the~simulation of 
charged\nobreakdash-particle reconstruction, 
the~track reconstruction efficiency
is corrected using 
a~sample of \mbox{$\decay{\jpsi}{\mumu}$}~decays 
in data~\cite{LHCb-DP-2013-002}.
Samples of 
the~\mbox{$\decay{\Bu}{\jpsi\Kstarp}$}~decays 
with 
\mbox{$\decay{\Kstarp}{\Kp\left(\decay{\piz}{\g\g}\right)}$}
are used to correct 
the~photon reconstruction efficiency  
in simulation~\cite{LHCb-PAPER-2012-022,
LHCb-PAPER-2012-053,
Govorkova:2015vqa,
Govorkova:2124605}.

\section{Event selection}
\label{sec:sel}

The \betapk candidates are reconstructed 
with $\etapr$~decays to
    either
\mbox{$\left(\decay{\Prho^0}{\pip\pim}\right)\g$}
or 
\mbox{$\left(\decay{\Peta}{\g\g}\right)\pip\pim$}
final states.
The~difficulty of reconstructing photons 
in the~\mbox{$\decay{\Peta}{\g\g}$}~decay 
leads to a~sample with fewer events.
The~\bpsik~normalisation decay
is reconstructed using 
the~\mbox{$\decay{\psitwos}{\jpsi\pip\pim}$}~decay. 
In~both signal and normalisation  
channels, the~\jpsi~meson 
is reconstructed in its decay to 
two muons.
As~explained in detail below, 
an~initial 
loose selection is 
applied for both 
signal and normalisation channels. 
Subsequently, for the~\betapk~candidates,
where the~background level is large, 
a~multivariate estimator
is used to select higher purity subset of candidates.
The~normalisation channel 
has a~high purity 
after 
the~initial selection, 
therefore no further selection 
steps are applied.

To reduce systematic uncertainties,  
the~initial 
selection criteria for both signal and normalisation channels 
are kept the~same whenever possible. 
The~selection criteria are chosen to be similar to those used in 
previous LHCb studies~\mbox{\cite{LHCb-PAPER-2012-022,
LHCb-PAPER-2012-053,
LHCb-PAPER-2013-024,
LHCb-PAPER-2014-008,
LHCb-PAPER-2014-056,
LHCb-PAPER-2021-003,
LHCb-PAPER-2021-047}}. 
The~muon, pion and kaon candidates are identified by combining 
information from the~Cherenkov~detectors,
calorimeters and muon detectors~\cite{LHCb-PROC-2011-008} 
associated
    with
the~reconstructed tracks. 
To~reduce the~combinatorial background, 
only tracks that 
are inconsistent with originating from any 
reconstructed PV in the~event are considered. 
The~transverse momentum of the~muon candidates is required to
be greater than 500\mevc and their momenta 
must
exceed 6\gevc. Pairs of oppositely\nobreakdash-charged muons 
consistent with originating from a~common vertex 
are combined to form \mbox{$\decay{\jpsi}{\mumu}$}~candidates.
The~reconstructed mass 
of the~muon pair is required 
to be between $3.056$~and $3.136\gevcc$.

Tracks that are consistent with 
the~pion or kaon hypotheses are required to 
have transverse momentum greater than~200\mevc. 
Photons are reconstructed from clusters 
in the~electromagnetic 
calorimeter, 
with transverse energy above~350\mev.
The~clusters must not be 
associated with reconstructed 
tracks~\cite{Terrier:691743,
LHCb-DP-2020-001}.
Photon identification is based on the~combined information 
from electromagnetic and hadronic calorimeters, 
scintillation pad, 
preshower detectors and the~tracking system. 

For the~reconstruction  of 
the~\etapetapipi~candidates, 
two photons are first combined 
to form an~$\Peta$~candidate. 
The~diphoton mass 
is restricted to lie within $\pm60\mevcc$ 
around 
the~known mass of the~$\Peta$~meson~\cite{PDG2022}. 
Each~$\Peta$~candidate is then combined with two 
oppositely\nobreakdash-charged pions 
to form an~\etapr~candidate. 
The~mass of the~combination is required to lie within 
$\pm45\mevcc$ around the~known mass of
the~\etapr~meson~\cite{PDG2022}.
For~an~\etapr~candidate  
reconstructed in 
the~\etaprhog~decay mode, 
the~$\Prho^0$~candidate is formed from
two 
oppositely\nobreakdash-charged 
pions. 
The~mass of this candidate is restricted to 
lie between 500 and 900\mevcc.
This asymmetric region around the~known mass of 
the~$\Prho^0$~meson~\cite{PDG2022}
takes into account the~shift 
of the~$\Prho^0$~line shape, 
due to the~electric\nobreakdash-dipole nature of
the~\mbox{$\decay{\etapr}
{\Prho^0\g}$}~transition~\cite{TASSO:1984xri,
Kolanoski:1984hu,
ARGUS:1990dcm,
CLEO:1991kmc,
CLEO:1992dty,
BESIII:2017kyd}.
A~photon 
is combined with 
the~$\Prho^0$~candidate
in order to form 
the~\etapr~candidate, 
whose mass 
is required to lie within 
$\pm30\mevcc$ of the~known \etapr~mass~\cite{PDG2022}.

Each selected \jpsi~candidate is combined with 
a~kaon track 
and either an~\etapr~candidate or 
two oppositely\nobreakdash-charged pions 
to form a~\Bp~candidate decaying 
into the~signal or normalisation modes, 
respectively. 
For the~\mbox{$\decay{\Bu}{\psitwos\Kp}$}~candidates
the~$\jpsi\pip\pim$~mass is required to be 
between $3.66$ and $3.71\gevcc$.
To~improve the~\Bp~meson mass resolution 
a~kinematic fit~\cite{Hulsbergen:2005pu} 
is performed,
which 
constrains 
the~masses of the~\jpsi, \etapr and \Peta~candidates 
to their known values~\cite{PDG2022}, 
and the~\Bp~candidates to originate from 
its associated PV.
The~decay time of 
the~\Bp~candidates is required to be greater 
than $100\mum/c$ to suppress 
the~large combinatorial 
background from tracks created in a~PV.

Further selection of
the~\mbox{$\decay{\Bu}{\jpsi\etapr\Kp}$}~decays 
is based 
on a~multivariate estimator,
in the~following referred to as 
the~multi\nobreakdash-layer 
perceptron\,({\sc{MLP}}) classifier.
The~classifier 
is based on an~artificial neural 
network algorithm~\cite{McCulloch,rosenblatt58}, 
configured with a~cross\nobreakdash-entropy cost 
estimator~\cite{Zhong:2011xm}.
It~reduces the~combinatorial background to a~low level 
while retaining a~high signal efficiency. 
Two~{\sc{MLP}}~classifiers are trained separately for 
the~two 
different \etapr~meson decay modes.
    The~list of variables
    used for classifiers includes
    the~\chisq of the~kinematic fit;
    transverse momenta of the~\etapr,
    kaon and pion candidates;
    pseudorapidities of pion and kaon candidates;
    transverse momentum of the~photon
    from the~\mbox{$\decay{\etapr}{\rhoz\g}$}~decay
    or miminal transverse momentum of
    photons from
    the~\mbox{$\decay{\etapr}{\left(\decay{\Peta}{\g\g}\right)\pip\pim}$}~decay;
    decay time  of the~\Bu~candidate; 
    variable related to the~quality of kaon
    identification~\cite{LHCb-PROC-2011-008,LHCb-DP-2012-003}
    and,
    for the~\mbox{$\decay{\etapr}{\rhoz\g}$}~decay,
        cosine of the~angle between momenta of the~\pip
        and \etapr~candidates in the~rest frame of the~\rhoz~candidate.
The~classifiers are trained using simulated samples 
of \betapk~decays as signal proxy, while  
the~\betapk~candidates from data with mass
above 5.35\gevcc are used to represent the background. 
The~\mbox{$\decay{\Bu}{\jpsi\left(\etaprhog\right)\Kp}$}
candidates  
with the~\jpsi\pip\pim\Kp mass
    consistent with
the~known mass of the~\Bp~meson 
are vetoed to avoid 
contamination from 
the~\mbox{$\decay{\Bp}{\jpsi\pip\pim\Kp}$}~decays 
with a~random photon added.

The~requirement on each of the~{\sc{MLP}} classifiers 
is chosen to maximise 
the~figure\nobreakdash-of\nobreakdash-merit 
defined as 
$S/\sqrt{B+S}$,
where $S$~represents 
the~expected signal yield,
and $B$~is the~expected background yield
within a~$\pm15\mevcc$ 
mass window centred
around the~known mass of 
the~$\Bu$~meson
    and corresponding to
    approximately
    three times the~mass resolution
    on both sides of the~peak.
The~background yield is calculated 
from fits to data,
    as described in Sec.~\ref{sec:sig},
while 
the~expected signal yield is estimated as $S=\varepsilon S_0$,
where $\varepsilon$ is the~efficiency of the~requirement 
on the~response of the~{\sc{MLP}}~classifier determined 
from simulation, 
and $S_0$~is the~signal yield obtained 
from the~fit to 
the~data,
with a~loose 
requirement
applied
on the~response of the~{\sc{MLP}}~classifier.\footnote{The~optimal requirement is found to be
  largely independent on the~choice
  of the~normalisation point.}
The~mass distributions 
for the~selected \mbox{$\decay{\Bu}
{\jpsi\etapr\Kp}$}~candidates 
are shown in Fig.~\ref{fig:signals}, 
where clear signal
peaks 
corresponding to \Bp~decays are seen in data for 
both \etapr decay modes.

\section{Signal yield determination}
\label{sec:sig}

\begin{figure}[t]
	\setlength{\unitlength}{1mm}
	\centering
	\begin{picture}(160,60)
	\put(  0, 0){\includegraphics*[width=80mm]{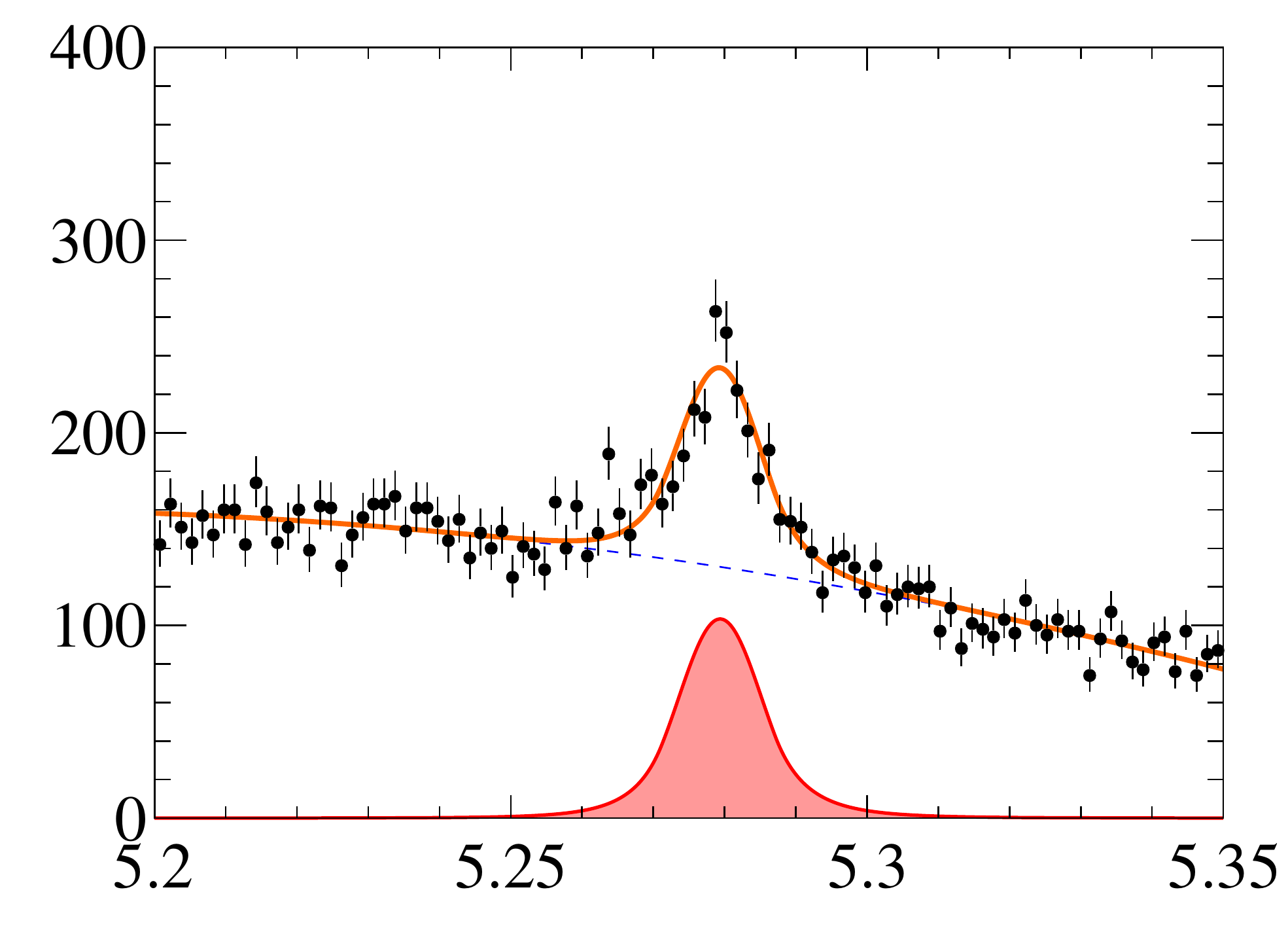}}
    \put( 80, 0){\includegraphics*[width=80mm]{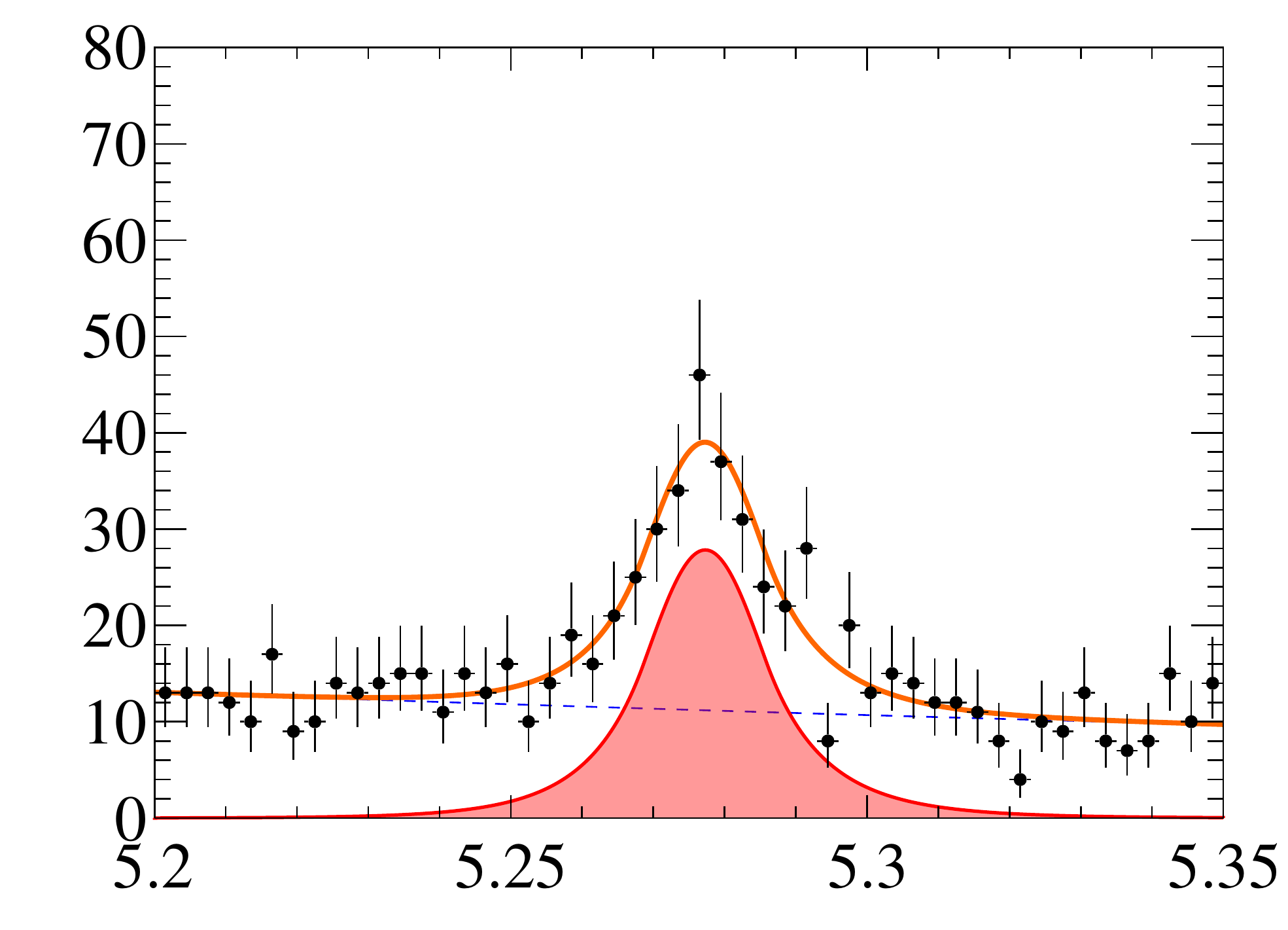}}
	\put( 37,-2){${m_{\jpsi\etapr\Kp}}$}
	\put( 62,-2){$\left[\!\gevcc\right]$}
	\put( -2,31){\rotatebox[]{90}{Candidates\,/\,(1.5\mevcc)}}
	\put(117,-1){${m_{\jpsi\etapr\Kp}}$}
	\put(142,-1){$\left[\!\gevcc\right]$}
	\put( 80,32){\rotatebox[]{90}{Candidates\,/\,(3\mevcc)}}
    \put( 55,37){$\decay{\etapr}{\Prho^0\g}$}
    \put(130,37){$\decay{\etapr}{\Peta\pip\pim}$}
 \put(12,43){\scriptsize
   $\begin{array}{cl}
    \!\bigplus\mkern-17mu\bullet & \text{Data}
   \\
   \begin{tikzpicture}[x=1mm,y=1mm]\filldraw[fill=red!35!white,draw=red,thick]  (0,0) rectangle (5,3);\end{tikzpicture}  
   & \decay{\Bu}{\jpsi\etapr\Kp} 
   \\
   {\color[RGB]{85,83,246}{\hdashrule[0.0ex][x]{5mm}{1.0pt}{2.0mm 0.3mm}}}
   & \text{Background}
   \\ 
   {\color[RGB]{255,153,51} {\rule{5mm}{2.0pt}}}
   & \text{Total}
   \end{array}$ }
\put(92,43){\scriptsize
   $\begin{array}{cl}
    \!\bigplus\mkern-17mu\bullet & \text{Data}
   \\
   \begin{tikzpicture}[x=1mm,y=1mm]\filldraw[fill=red!35!white,draw=red,thick]  (0,0) rectangle (5,3);\end{tikzpicture}  
   & \decay{\Bu}{\jpsi\etapr\Kp} 
   \\
   {\color[RGB]{85,83,246}{\hdashrule[0.0ex][x]{5mm}{1.0pt}{2.0mm 0.3mm}}}
   & \text{Background}
   \\ 
   {\color[RGB]{255,153,51} {\rule{5mm}{2.0pt}}}
   & \text{Total}
   \end{array}$ } 
   \put( 58,45){$\begin{array}{l}\lhcb\\ 9\invfb\end{array}$}
   \put(138,45){$\begin{array}{l}\lhcb\\ 9\invfb\end{array}$}
	\end{picture}
	\caption { \small 
     Mass distributions 
     for selected \betapk~candidates 
     with $\etapr$~decays 
     to (left)~$\rhoz\g$
     and (right)~$\Peta\pip\pim$~final states. 
         The~resulting fit functions
         are overlaid with data distributions.
        }
	\label{fig:signals}
\end{figure}
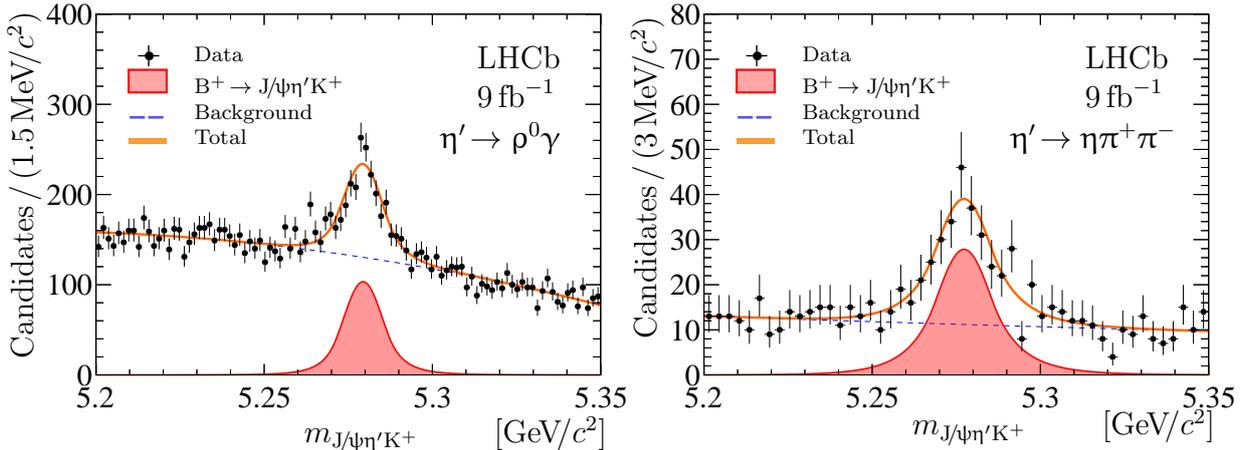

The signal yields are determined using an~extended unbinned 
maximum\nobreakdash-likelihood fit
    to the~\mbox{$\jpsi\etapr\Kp$}~mass distributions
with a~two\nobreakdash-component function.
The~signal component for both cases is modelled 
by a~modified Gaussian function
    that combines a~Gaussian core
with power\nobreakdash-law tails 
on both sides of the~distribution~\cite{Skwarnicki:1986xj,
LHCb-PAPER-2011-013}. 
The~background component 
is parameterised by a~second\nobreakdash-order 
positive polynomial function~\cite{karlin1953geometry}
for
the~\etaprhog~decay mode 
and by an~exponential function in 
case of 
the~\etapetapipi~decay mode.
    The~parameters of the~detector
    resolution
    function are taken from simulation,
    and the~width of the~Gaussian function is
further 
corrected by a~scale factor, $s_{\Bu}$,
that accounts for a~small discrepancy between data and 
simulation~\cite{LHCb-PAPER-2020-009,
LHCb-PAPER-2020-035,
LHCb-PAPER-2021-034,
LHCb-PAPER-2021-047,
LHCb-PAPER-2022-025}. 
To~account for the~uncertainty in 
the~tail parameters and resolution, 
the~fit is performed simultaneously 
for data and simulated samples, 
sharing the~same tail parameters, 
and allowing the~correction factor $s_{\Bu}$~to vary.
    The~resulting fit functions are overlaid
    with data distributions
in Fig.~\ref{fig:signals}
and the~signal yields are found to be 
\begin{subequations}
\begin{eqnarray}
 \left. N_{\decay{\Bu}{\jpsi\etapr\Kp}}
 \right|_{\decay{\etapr}{\Prho^0\g}} 
     & = & 
    \left(1.11\phantom{0}\pm0.11\phantom{0}\right)\times 10^3\,, \label{eq:ns_rhog}
  \\
 \left. 
  N_{\decay{\Bu}{\jpsi\etapr\Kp}}
  \right|_{ \decay{\etapr}{\Peta\pip\pim} }
   & = &  
 \left( 0.228 \pm 0.028 \right) \times 10^3 \,,
   \label{eq:ns_etapipi}
\end{eqnarray}
\end{subequations}
where the~uncertainties are statistical only. 
The~resolution correction factors $s_{\Bu}$~are found to be 
$1.08 \pm 0.12$ 
and $1.03 \pm 0.16$ 
for the~\etaprhog~and 
\etapetapipi~samples, respectively.  
In both cases, 
the~statistical significance of the 
\betapk~signal is 
calculated using Wilks' theorem~\cite{Wilks:1938dza}
and found 
to exceed  $17$ and $12$~standard deviations
for the~\etaprhog~and 
\etapetapipi~samples, respectively.  
However, as the~signal yield is much lower 
for the~\etapr~meson 
decays to the~$\Peta\pip\pim$~final state, 
all~subsequent studies are performed using only 
the~\etaprhog~decay mode.

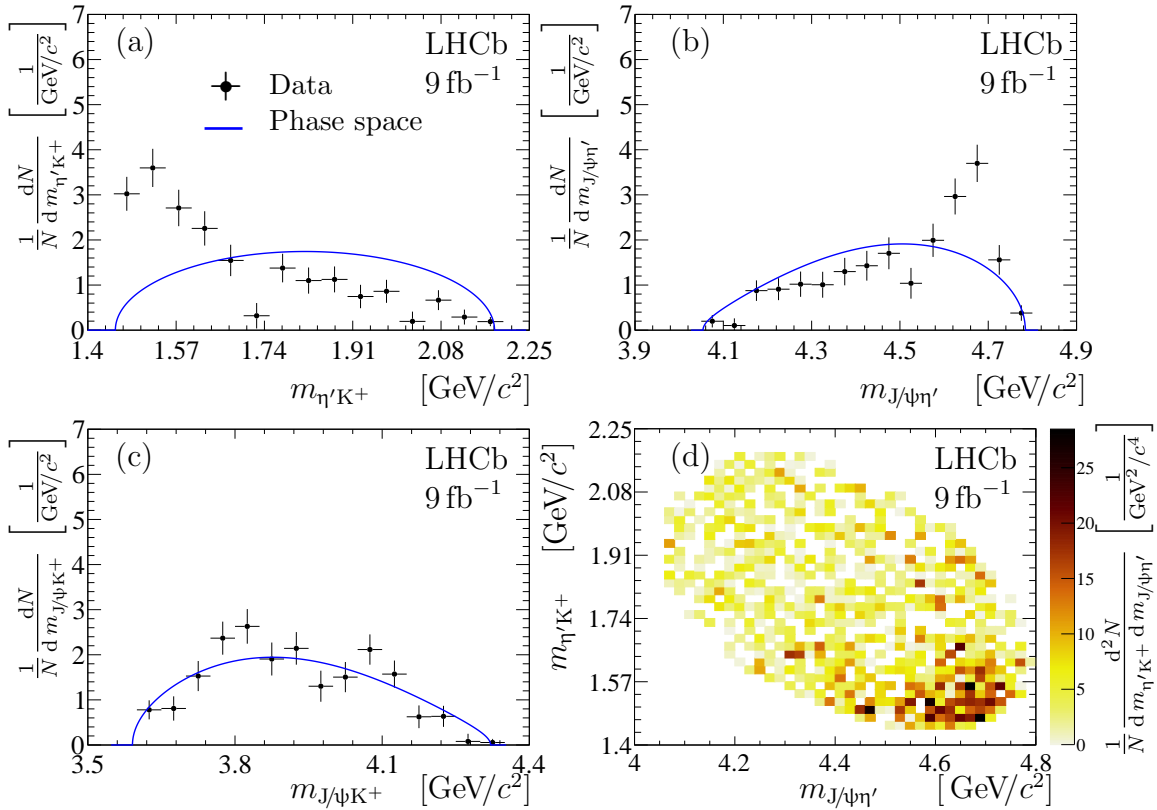
\begin{figure}[t]
	\setlength{\unitlength}{1mm}
	\centering
	\begin{picture}(150,120)
	\put( 0,55){\includegraphics*[width=70mm]{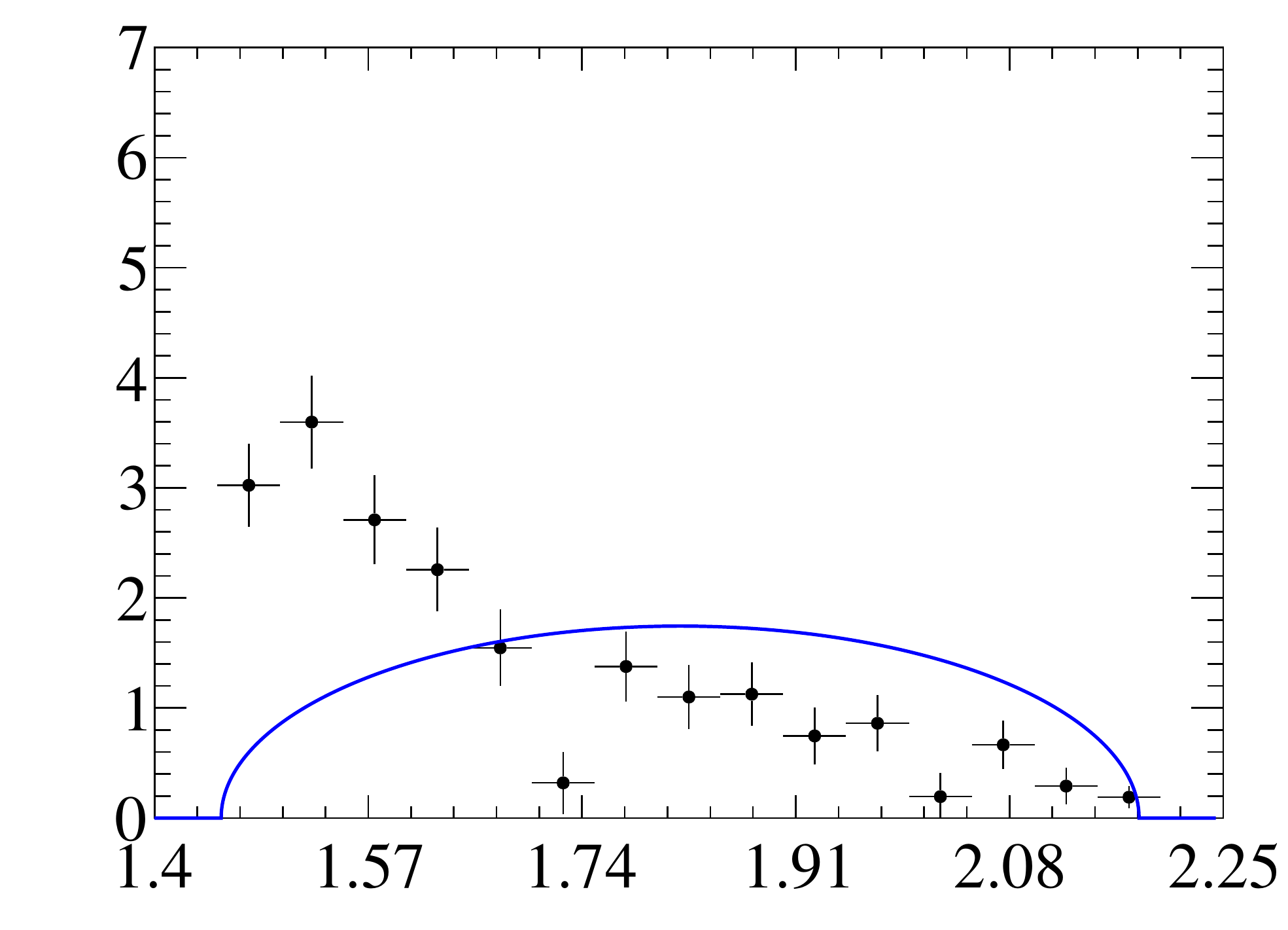}}
	\put(72,55){\includegraphics*[width=70mm]{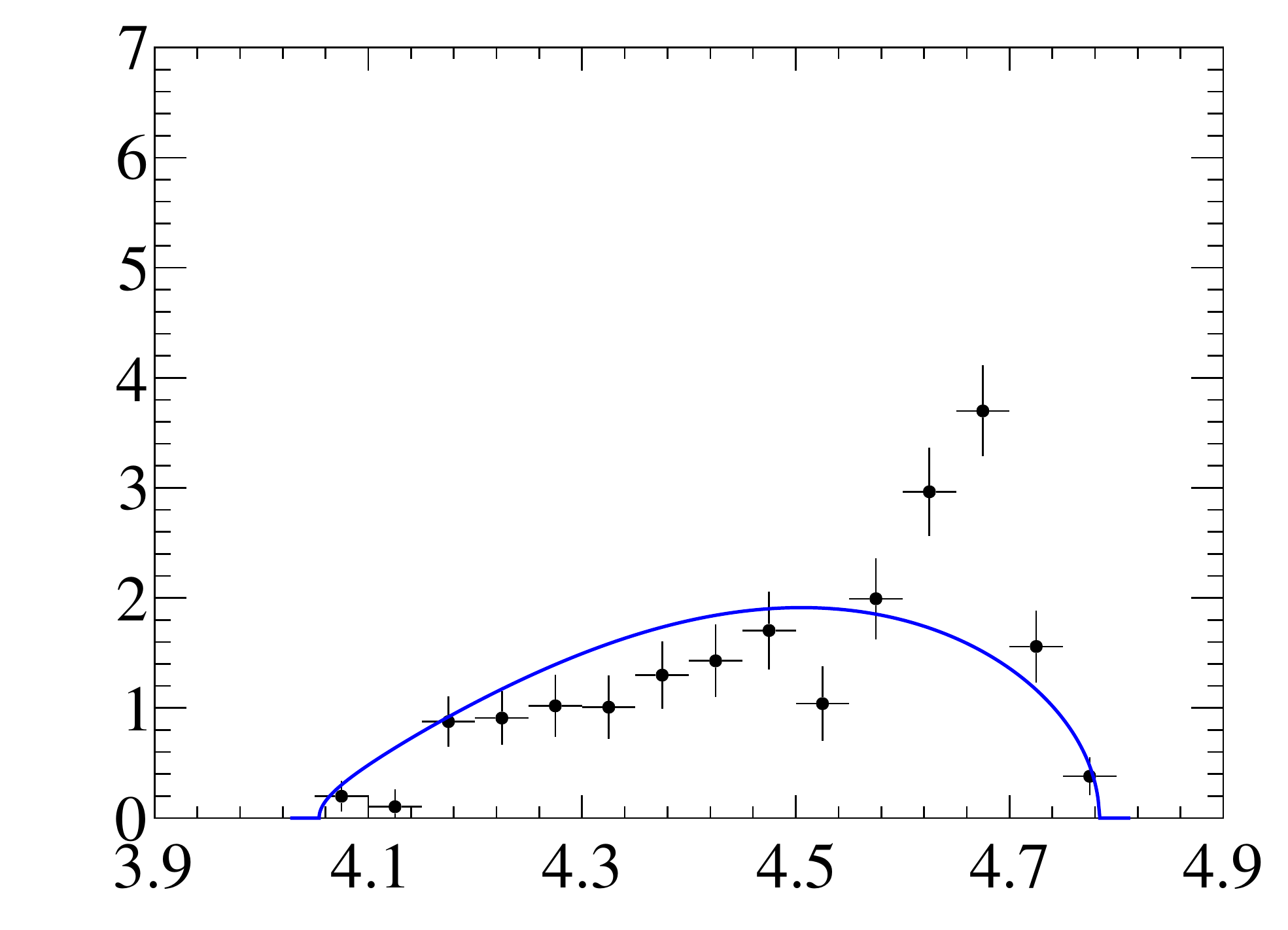}}
	\put( 0, 0){\includegraphics*[width=70mm]{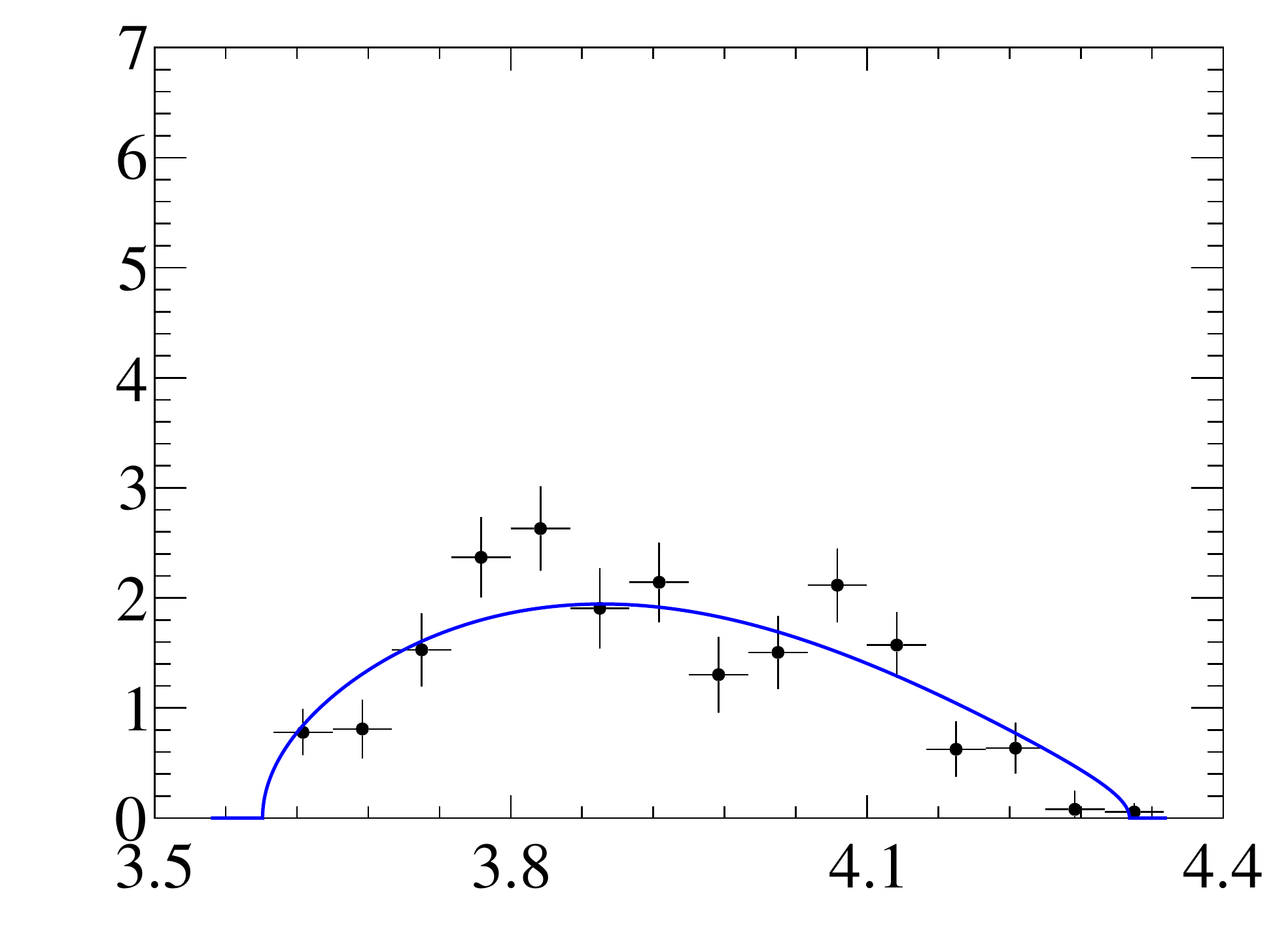}}
	\put(72, 0){\includegraphics*[width=70mm]{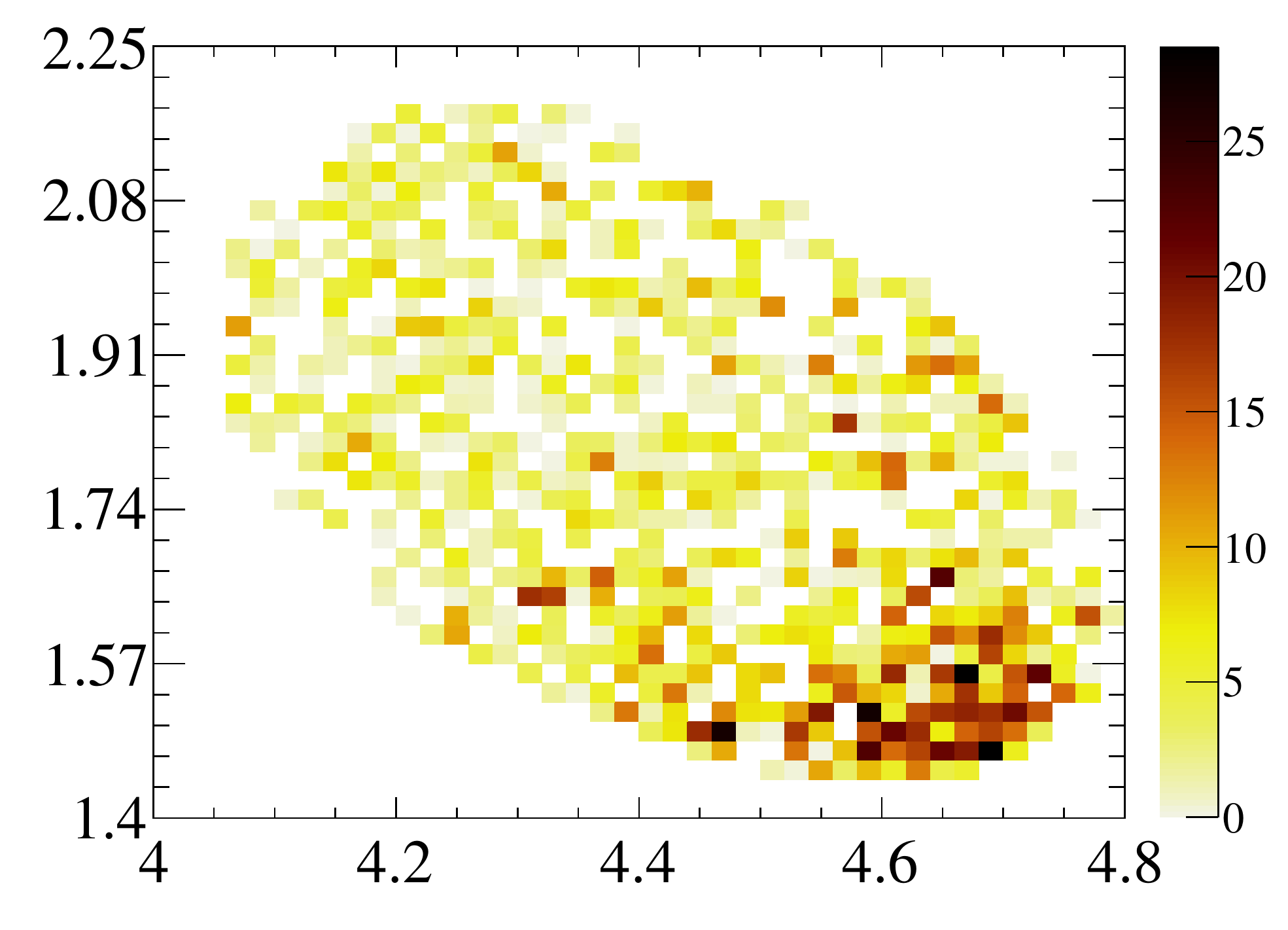}}
	\put( 35,52){${m_{\etapr\Kp}}$}
	\put(110,52){${m_{\jpsi\etapr}}$}
	\put( 35,-1){${m_{\jpsi\Kp}}$}
        \put(102,-1){${m_{\jpsi\etapr}}$}
	\put( 52, 52){$\left[\!\gevcc\right]$}
	\put(125, 52){$\left[\!\gevcc\right]$}
	\put( 52, -1){$\left[\!\gevcc\right]$}
        \put(120, -1){$\left[\gevcc\right]$}
	\put( -2, 86){\rotatebox[]{90}
        {$\frac{1}{N}\frac{\mathrm{d}N}
        {\mathrm{d}\,m_{\etapr\Kp}}\,\left[\frac{1}{\!\gevcc}\right]$}}
  	\put( 68, 86){\rotatebox[]{90}
        {$\frac{1}{N}\frac{\mathrm{d}N}
        {\mathrm{d}\,m_{\jpsi\etapr}}\,\left[\frac{1}
        {\!\gevcc}\right]$}}
        \put( -2, 30){\rotatebox[]{90}
        {$\frac{1}{N}\frac{\mathrm{d}N}
        {\mathrm{d}\,m_{\jpsi\Kp}}\,\left[\frac{1}
        {\!\gevcc}\right]$}}
        \put( 68, 30){\rotatebox[]{90}
        {$m_{\etapr\Kp}~~~\left[\!\gevcc\right]$}}
       \put(141, 26){\rotatebox[]{90}
        {$\frac{1}{N}\frac{\mathrm{d}^2N}
        {\mathrm{d}\,m_{\etapr\Kp}\,
        \mathrm{d}\,m_{\jpsi\etapr}}
        \,\left[\frac{1}
        {\!\gev^2/c^4}\right]$}}
        
    \put(124,95){$\begin{array}{l}\lhcb\\ 9\invfb\end{array}$}
    \put( 51,95){$\begin{array}{l}\lhcb\\ 9\invfb\end{array}$}
    \put(118,40){$\begin{array}{l}\lhcb\\ 9\invfb\end{array}$}
    \put( 51,40){$\begin{array}{l}\lhcb\\ 9\invfb\end{array}$}
    \put(12,  98){(a)}
    \put(85,  98){(b)}
    \put(12,  43){(c)}
    \put(85,  43){(d)}
    \put(22,90){\small
   $\begin{array}{cl}
    \!\bigplus\mkern-18mu{\scriptsize{\bullet}} & \text{Data}
   \\
   {\color[RGB]{0,0,255} {\rule{5mm}{1.0pt}}}
   & \text{Phase~space}
   \end{array}$ }
	\end{picture}
	\caption { \small 
        Normalised background\protect\nobreakdash-subtracted 
        (a)~$\etapk$,
        (b)~$\psietap$,
        (c)~$\psik$~mass spectra 
        and 
        (d)~two\protect\nobreakdash-dimensional mass distribution of
        $\etapr\Kp$ vs $\jpsi\etapr$
        from 
        the~\mbox{$\decay{\Bu}{\jpsi\etapr\Kp}$}~decays.
        Superimposed curves are the expectations 
        from a~phase\protect\nobreakdash-space model.
        }
	\label{fig:splots}
\end{figure}

The~background\nobreakdash-subtracted  
$\psietap$, 
$\etapk$,
and $\psik$~mass spectra 
from 
the~\mbox{$\decay{\Bu}
{\jpsi\left(\decay{\etapr}{\Prho^0\g}\right)\Kp}$}~decays
are shown in Figs.~\ref{fig:splots}(a-c),
where 
the~\sPlot~technique~\cite{Pivk:2004ty} 
based on the~fit results
is used for background subtraction. 
The~$\psietap$, 
$\etapk$,
and $\psik$~masses are calculated 
using a~kinematic fit 
with \jpsi, \etapr and \Bp mass constraints 
and a~PV constraint applied~\cite{Hulsbergen:2005pu}. 
While for the~$\jpsi\Kp$~mass 
the~distribution largely agrees with 
the~shape expected from the~phase\nobreakdash-space model, 
for the~low\nobreakdash-mass 
region of the~$\etapk$~mass spectrum 
and the~high\nobreakdash-mass 
region 
of the~$\psietap$~mass spectrum 
a~striking difference from the~phase\nobreakdash-space model 
is observed.
These differences 
are 
potentially 
due to 
contributions 
from decays via intermediate heavy 
excited strange mesons,
such as 
$\PK^{*}_{0}(1430)^{+}$, 
$\PK^{*}_{2}(1430)^{+}$ or 
$\PK^{*}(1680)^{+}$~mesons, 
decaying into 
the~$\etapr\Kp$~final state.
The~decays 
of the~\Bp~mesons into a~\jpsi~meson 
and heavy excited strange mesons 
have been
studied 
in 
Refs.~\cite{LHCb-PAPER-2016-018,
LHCb-PAPER-2016-019,
LHCb-PAPER-2020-044}.
    The~decays via intermediate excited kaons
    also contribute to
the~higher mass region of the~$\jpsi\etapr$~mass spectrum,
as shown in Fig.~\ref{fig:splots}(d).
The~$\jpsi\etapr$~mass region below~$4.7\gevcc$
is explicitly inspected for 
possible contributions 
from decays via excited
    charmonium or charmonium\nobreakdash-like
states into the~$\jpsi\etapr$~final state. 
Fits~to the~background\nobreakdash-subtracted 
\psietap~mass distribution are performed 
in individual mass windows,
corresponding to 
the~well\nobreakdash-established 
$\Ppsi(4160)$, $\Ppsi(4230)$,
    $\Ppsi(4360)$,
$\Ppsi(4415)$ 
and~$\Ppsi(4430)$~resonances~\cite{PDG2022}. 
For each fit, the~resonance shape is parameterised
with a~relativistic Breit\nobreakdash--Wigner function 
convoluted with a~mass 
resolution function.
The~non\nobreakdash-resonant contribution is modelled 
by a~first order
    positive
polynomial function.
The~known masses and widths of the resonances~\cite{PDG2022} 
are introduced in the~fits as Gaussian constraints 
on the~corresponding parameters.
    The~resolution function
    is modelled by the~modified Gaussian
    function with parameters 
    obtained using 
    simulation as a~function of 
    the~\psietap mass.
No~statistically significant signals 
are observed 
for the~\mbox{$\decay{\Bp}{\jpsi\etapr\Kp}$}~decays via 
intermediate 
resonances, listed above.
    To~probe the contribution
    of the~resonances with higher masses,
    more advanced fit techniques
    accounnting for complicated background shape
    and  the~distortion of the~signal Breit\nobreakdash--Wigner
    shape are required.

For the~determination of the~resonant structure
of the~\mbox{$\decay{\Bu}{\jpsi\etapr\Kp}$}~decay,
a~full amplitude analysis, 
similar to those used in 
Refs.~\cite{LHCb-PAPER-2016-019,
LHCb-PAPER-2020-044}, 
is required.
Large signal yields and 
low background levels 
are important prerequisites 
for such analysis. 
The~relatively large level of 
combinatorial background 
for 
the~\mbox{$\decay{\Bu}{\jpsi\etapr\Kp}$}~signal 
decays with the~\etapr~meson 
reconstructed via 
the~$\etaprhog$~decay mode, 
makes it difficult to carry out an~amplitude analysis.
With a~larger data sample, 
expected from 
future 
data\nobreakdash-taking periods,  
it will be possible to 
perform the~full amplitude 
analysis using 
the~\mbox{$\decay{\Bu}{\jpsi\etapr\Kp}$}~signal 
decays with the~\etapr~meson 
reconstructed via 
the~$\etapetapipi$~decay mode,
where the~combinatorial background 
is smaller.

\section{Normalisation channel}
\label{sec:norm}

\begin{figure}[t]
	\setlength{\unitlength}{1mm}
	\centering
	\begin{picture}(160,60)
	\put(0 , 0){\includegraphics*[width=80mm]{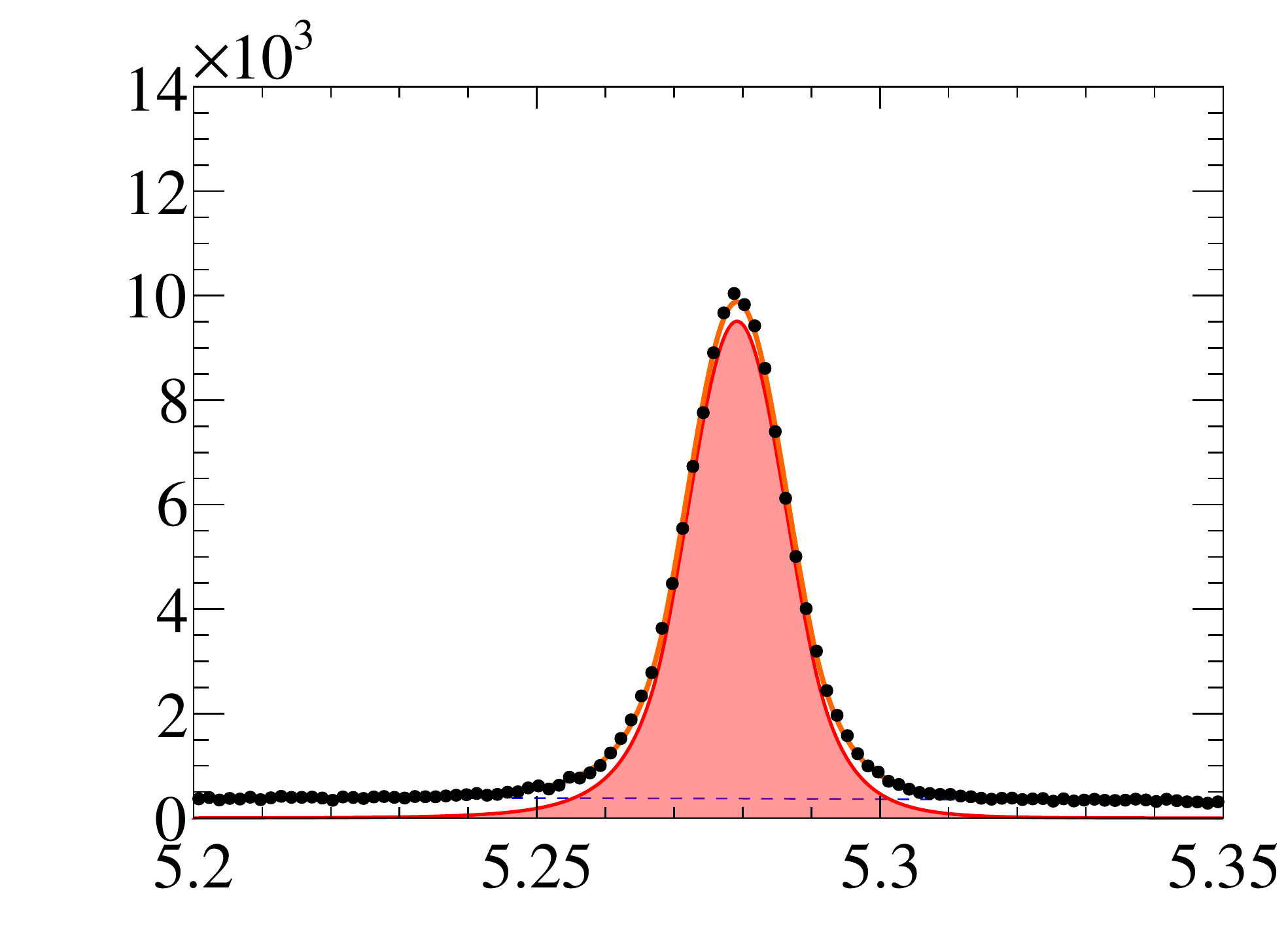}}
	\put(80, 0){\includegraphics*[width=80mm]{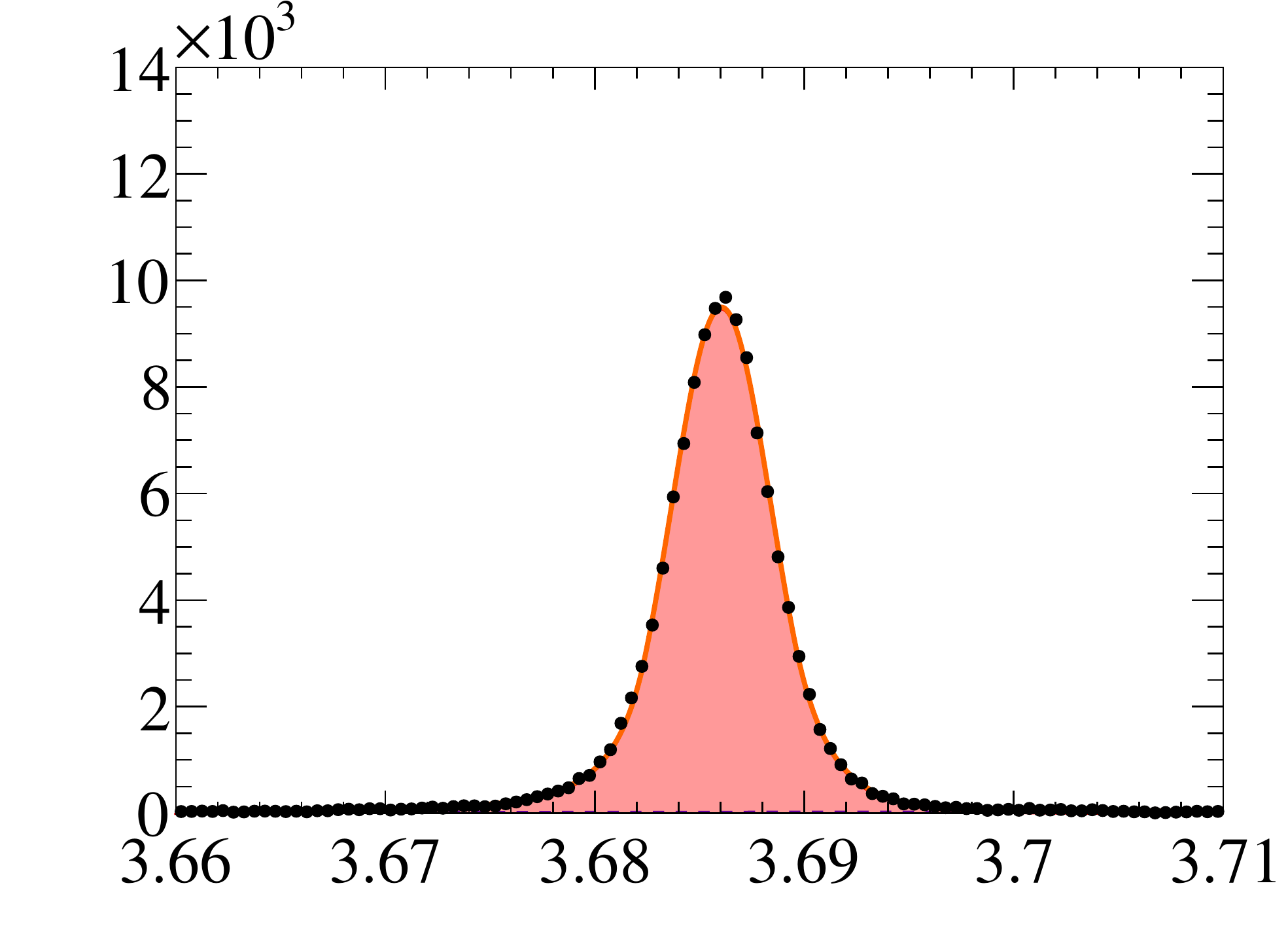}}
	\put( 35,-2){${m_{\psipipi\Kp}}$}
	\put(115,-2){${m_{\psipipi}}$}
	
        \put( 60,-2){$\left[\!\gevcc\right]$}
	\put(140,-2){$\left[\!\gevcc\right]$}
	\put( 2,29){\rotatebox[]{90}{Candidates\,/\,(1.5\mevcc)}}
	\put(82,34){\rotatebox[]{90}{Yield\,/\,(0.5\mevcc)}}
    \put( 59,42){$\begin{array}{l}\lhcb\\ 9\invfb\end{array}$}
    \put(139,42){$\begin{array}{l}\lhcb\\ 9\invfb\end{array}$}
 \put(15,42){\scriptsize
   $\begin{array}{cl}
    \!\bigplus\mkern-17mu\bullet & \text{Data} 
   \\
   \begin{tikzpicture}[x=1mm,y=1mm]\filldraw[fill=red!35!white,draw=red,thick]  (0,0) rectangle (5,3);\end{tikzpicture}  
   & \decay{\Bu}{\jpsi\pip\pim\Kp} 
   \\
   {\color[RGB]{85,83,246}{\hdashrule[0.0ex][x]{5mm}{1.0pt}{2.0mm 0.3mm}}}
   & \text{Background}
   \\ 
   {\color[RGB]{255,153,51} {\rule{5mm}{2.0pt}}}
   & \text{Total}
   \end{array}$ }
\put(94,42){\scriptsize
   $\begin{array}{cl}
     \!\bigplus\mkern-17mu\bullet & \text{Data}
   \\
   \begin{tikzpicture}[x=1mm,y=1mm]\filldraw[fill=red!35!white,draw=red,thick]  (0,0) rectangle (5,3);\end{tikzpicture}  
   & 
   \decay{\Bu}{\psitwos\Kp}
   \\
   {\color[RGB]{85,83,246}{\hdashrule[0.0ex][x]{5mm}{1.0pt}{2.0mm 0.3mm}}}
   & \text{Non-res.} 
   \\ 
   {\color[RGB]{255,153,51} {\rule{5mm}{2.0pt}}}
   & \text{Total}
   \end{array}$ }

	\end{picture}
	\caption { \small
    Left:~mass distributions 
    for selected \mbox{$\decay{\Bu}{\left(\decay{\psitwos}{\jpsi\pip\pim}\right)\Kp}$} decays, with 
    $\jpsi\pip\pim$~mass between 3.66~and~3.71\gevcc.
    Right:~background-subtracted  
    $\jpsi\pip\pim$~mass distribution 
    from selected \Bu~decays.  
    The~resulting fit functions
          are overlaid
          with data distributions.
    }
	\label{fig:norm_intermediates}
\end{figure}
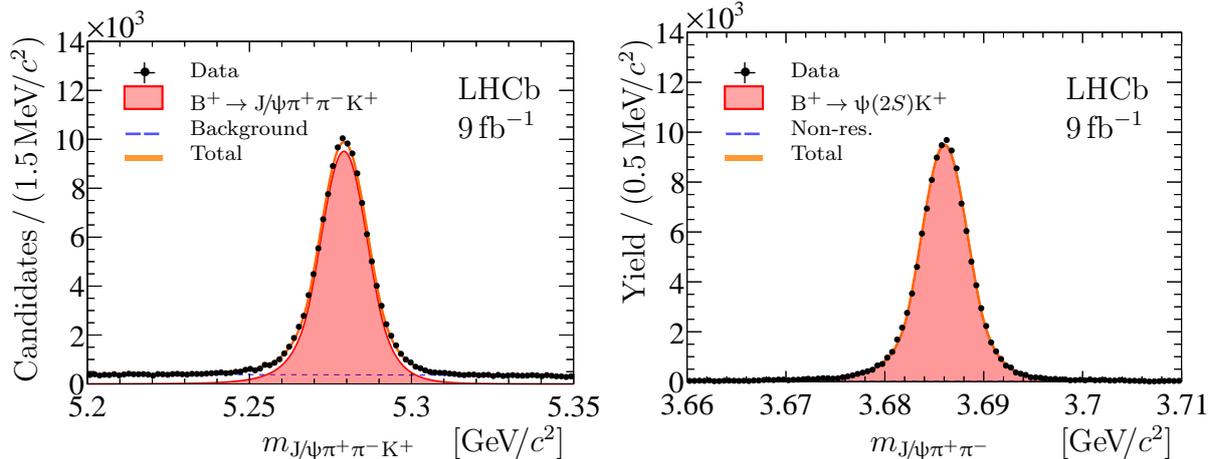

The \jpsi\pip\pim\Kp mass distribution 
for selected $\decay{\Bu}{\left(\decay{\psitwos}
{\jpsi\pip\pim}\right)\Kp}$~candidates
with $\jpsi\pip\pim$~mass between 3.66~and 3.71\gevcc 
is shown in Fig.~\ref{fig:norm_intermediates}(left). 
An~extended unbinned 
maximum\nobreakdash-likelihood 
fit is performed,
where 
the~signal component
is modelled by 
the~modified Gaussian function
and the~background 
is described by 
a~first order
positive polynomial
function. 
The~result of this~fit is used to 
obtain the~background\nobreakdash-subtracted
$\jpsi\pip\pim$~mass distribution 
from \mbox{$\decay{\Bu}{\jpsi\pip\pim\Kp}$}~decays. 
This~distribution is shown in
Fig.~\ref{fig:norm_intermediates}(right). 
The~yield of 
the~\mbox{$\decay{\Bu}
{\left(\decay{\psitwos}
{\jpsi\pip\pim}\right)\Kp}$}~signal
candidates is determined 
using an~unbinned fit to this distribution with 
a~two\nobreakdash-component function.
The~component corresponding 
to the~\mbox{$\decay{\Bu}{\psitwos\Kp}$}~decays
is parameterised 
by
the~modified Gaussian function.
The~component describing
the~\mbox{$\decay{\Bu}{\jpsi\pip\pim\Kp}$}~decays
without intermediate \psitwos~state 
is modelled by
a~phase\nobreakdash-space 
function,\footnote{
The~phase\nobreakdash-space mass distribution of 
a~$k$\nobreakdash-body combination of
particles from an~$n$\nobreakdash-body decay 
is approximated by 
$\Phi_{k,n}(x) \propto 
x_{\ast}^{ (3k-5)/2}\left(1-x_{\ast}\right)^{3(n-k)/2-1}$,
where 
$x_{\ast}\equiv 
( x-x_{\mathrm{min}} ) /(x_{\mathrm{max}}-x_{\mathrm{min}})$, 
and 
$x_{\mathrm{min}}$, $x_{\mathrm{max}}$ denote 
the~minimal   and maximal values of $x$, 
respectively~\cite{Byckling}.
Here, $k=3$ and $n=4$ are used.
} 
modified by 
a~first order positive
polynomial function.
The~tail and resolution parameters 
for the~signal component 
are taken from simulation
with the~resolution 
further corrected 
by a~scale factor, $s_{\psitwos}$, 
that accounts for a~small discrepancy 
between data and simulation~\cite{LHCb-PAPER-2020-008,
  LHCb-PAPER-2020-009}.
    The~fit is performed
      simultaneously to
      data and simulated samples,
      as for the~signal~mode
      described in Sec.~\ref{sec:sig}.
From this fit the~number of 
\mbox{$\decay{\Bu}{\psitwos\Kp}$}~decays 
is found to be
\begin{equation}\label{eq:nn}
    N_{\decay{\Bu}{\psitwos\Kp}} = \left( 121.40 \pm 0.14\right)\times 10^{3}\,, 
\end{equation}
where the uncertainty is statistical only. 
The~correction factor 
$s_{\psitwos}$~is found to be $1.057 \pm 0.008$, which is
in good~agreement with results from 
Refs.~\cite{LHCb-PAPER-2020-008,
LHCb-PAPER-2020-009,
LHCb-PAPER-2020-035,
LHCb-PAPER-2021-034,
LHCb-PAPER-2022-025,
LHCb-PAPER-2022-049}. 

\section{Branching 
fraction ratio computation}
\label{sec:bf}

The~ratio of the~branching fractions $\mathcal{R}$, 
defined in Eq.~\eqref{eq:ratio},
is calculated as 
\begin{equation*}
\mathcal{R}
= 
\dfrac{N_{\decay{\Bu}{\jpsi\etapr\Kp}}}
      {N_{\decay{\Bu}{\psitwos\Kp}}}
\times
\dfrac{\BF(\psitwos\to\jpsi\pip\pim)}{\BF(\etaprhog)}
\times 
\dfrac{\epsilon_{\decay{\Bu}{\psitwos\Kp}}}
      {\epsilon_{\decay{\Bu}{\jpsi\etapr\Kp}}} 
\,,
\end{equation*}
where 
$N_{\decay{\Bu}{\jpsi\etapr\Kp}}$
and 
$N_{\decay{\Bu}{\psitwos\Kp}}$
are the~yields 
from Eqs.~\eqref{eq:ns_rhog} and~\eqref{eq:nn},  
and 
$\epsilon_{\decay{\Bu}{\jpsi\etapr\Kp}}$ 
and 
$\epsilon_{\decay{\Bu}{\psitwos\Kp}}$ 
are the~efficiencies
to reconstruct the~observed final states.
 The~efficiencies are the products of detector acceptance, 
reconstruction, selection 
 and trigger efficiencies, 
and 
 are calculated using simulated samples, 
 calibrated to match the data as 
 described in Sec.~\ref{sec:detector}.
     The~branching fractions for
    the~\mbox{\decay{\psitwos}{\jpsi\pip\pim}}
    and \mbox{\decay{\etapr}{\rhoz\g}}~decays,
    \mbox{$\BR\left(\decay{\psitwos}{\jpsi\pip\pim}\right)= \left( 34.68 \pm 0.30\right)\%$}
    and
    \mbox{$\BR\left(\decay{\etapr}{\rhoz\g}\right)= \left( 29.4 \pm 0.4 \right)\%$},
    are taken from Ref.~\cite{PDG2022}.
The~ratio of branching fractions $\mathcal{R}$ is found to be
\begin{equation*}\label{eq:R}
\dfrac{\BF(\betapk)}{\BF(\bpsik)} = 
\left(4.91\pm0.47\right)\times10^{-2},
\end{equation*}
where the uncertainty is statistical only. 

As a~cross\nobreakdash-check, the~ratio 
$\mathcal{R}$~is also calculated for 
the~\mbox{$\decay{\etapr}{\Peta\pip\pim}$},
using the~signal yields 
from Eqs.~\eqref{eq:ns_etapipi} 
and~\eqref{eq:nn}.
The~ratio 
of branching fractions
$\mathcal{R}$ is found to be 
\mbox{$\left(4.98\pm0.61\right)\times10^{-2}$}, 
where the~uncertainty is statistical only. 
This value  is in 
good agreement with 
that 
calculated for the~\etaprhog~case.

\section{Systematic uncertainties}

The~signal and normalisation channels share 
the~same set of final\nobreakdash-state charged particles, 
and the~same trigger and preselection requirements
are applied to both.
This allows 
many 
systematic uncertainties 
to cancel in the~ratio~$\mathcal{R}$. 
The~remaining nonnegligible 
contributions are listed in Table~\ref{tab:syst}.

Several systematic uncertainties are
associated 
    with
the~corrections applied to the~simulation. 
The~finite size of
the~$\decay{\Bp}{\jpsi\Kp}$ signal sample 
used for correction 
of the~simulated
transverse momentum and rapidity spectra 
of \Bp~mesons 
induces 
an~uncertainty
on the~\Bu~meson \pt and $y$~spectra.
In~turn, 
this uncertainty 
induces 
small changes in the~ratio of efficiencies. 
The~corresponding spread of these changes amounts to $0.1\%$ 
and is taken as the~systematic uncertainty
related to the~\Bu~meson kinematic.

The decay model corrections for 
the~\betapk~decay are obtained 
using 
the~algorithm described in Ref.~\cite{Rogozhnikov:2016bdp}.
The~systematic uncertainty related to the~correction 
method is estimated by varying 
the~configuration parameters of the~algorithm.
The~largest deviation of the~efficiency value 
from the~baseline tuning is found to be $1.1\%$, 
which is assigned as systematic uncertainty 
associated with the~\Bu~decay model. 

There are residual differences in 
the~reconstruction efficiency 
of charged\nobreakdash-particle tracks that 
do not cancel completely in the~ratio 
of total efficiencies 
given 
the~slightly~different kinematic distributions 
of the~final\nobreakdash-state particles.
The~track\nobreakdash-finding efficiencies 
obtained from simulated samples are corrected using
calibration channels~\cite{LHCb-DP-2013-002}.
The~uncertainties related to~the~efficiency 
correction factors are propagated to the~ratios of 
the~total efficiencies using pseudoexperiments, 
and are found to be 0.7\%. 
This value is taken as the~systematic uncertainty 
due to the~tracking efficiency calibration.

Differences
    in
the~photon reconstruction efficiencies 
between data and simulation 
are studied using a~large sample of 
\mbox{$\decay{\Bu}{\jpsi\Kstarp}$}~decays, 
reconstructed using 
the~\mbox{$\decay{\Kstarp}{\Kp\left(\decay{\piz}{\g\g}\right)}$}
decay mode~\cite{LHCb-PAPER-2012-022,
LHCb-PAPER-2012-053,
Govorkova:2015vqa,
Govorkova:2124605}.
The~uncertainty due to  
the~finite size of the~sample
is propagated to the~ratio of 
the~total efficiencies 
using pseudoexperiments
and is found
to be less 
than~1.0\%.
The~uncertainty due to the~accuracy 
of the~\mbox{$\decay{\Bu}{\jpsi\Kstarp}$}~branching 
fraction~\cite{PDG2022}
is 3.5\%.
These two values are added in quadrature to obtain 
a~systematic uncertainty related to 
the~photon reconstruction of~3.6\%.

The~kaon identification variable used for the~{\sc{MLP}}~estimator 
is drawn from calibration data samples
and has a~dependence on the~particle
kinematics and track multiplicity. 
Systematic uncertainties in this procedure arise from 
the~limited
size 
of both the~simulation 
and calibration samples, 
and the~modelling of the~particle identification variable. 
The~limitations due to
the~size of the~simulation 
and calibration samples 
are evaluated by 
using bootstrapping
techniques~\cite{efron:1979,efron:1993}, 
creating  
multiple samples
and repeating the~procedure 
for
    each of these.
The~impact of potential mismodelling of the~kaon identification  
variable is evaluated by describing 
the~corresponding distributions using 
density estimates with different kernel 
widths~\cite{LHCb-DP-2018-001,Poluektov:2014rxa}. 
For~each of these cases, alternative efficiency maps are 
produced to determine the~associated uncertainties. 
A~systematic uncertainty of 2.8\% is assigned from 
the~observed differences with alternative efficiency maps. 

A~systematic uncertainty  related 
to the~knowledge of the~trigger efficiencies 
has been previously studied using large 
samples of \mbox{$\decay{\Bu}
{\left(\decay{\jpsi}{\mumu}\right)\Kp}$} 
and \mbox{$\decay{\Bu}
{\left(\decay{\psitwos}{\mumu}\right)\Kp}$}~decays
by comparing the 
ratios of the trigger efficiencies in data 
and simulation~\cite{LHCb-PAPER-2012-010}.
Based on this comparison, a~relative uncertainty 
of 1.1\% is assigned.

The~remaining inconsistency between data and simulation, 
not covered by the~corrections discussed 
in Sec.~\ref{sec:detector}, 
is estimated by varying 
the~requirement on the~response
of the~{\sc{MLP}} classifier
in ranges that lead to
changes in the~measured signal yields 
as large as $\pm 20\%$. 
The~resulting difference 
in the~data\nobreakdash-simulation efficiency ratio 
is found to be~$3.0\%$. 

A~different class of systematic uncertainties 
directly affects the~fit itself, namely 
uncertainties associated 
    with
the~fit 
models used to describe the~\jpsi\etapr\Kp, 
\psitwos\Kp and \jpsi\pip\pim spectra. 
The~systematic uncertainty is accounted for 
by fits with alternative models. 
The~list of alternative models to describe 
signal \Bp and \psitwos components includes 
a~modified Apollonios function~\cite{MartinezSantos:2013ltf},
which has exponential instead 
of power\nobreakdash-law tails, 
a~generalised 
Student's $t$-distribution~\cite{Student,Jackman} 
and 
a~modified Novosibirsk function~\cite{BaBar:2011qjw}.
For~the~combinatorial 
background for 
both the~signal and 
the~normalisation channels, 
a~third order polynomial function 
and an~exponential 
function multiplied by 
a~first order polynomial
function
are chosen 
as alternative models.
For~each fit only 
one component\,(either signal or background) 
is replaced at a~time, and 
the~same fit function is used for 
both signal and normalisation channel. 
The~largest deviation of the~ratio of signal
yields is found to be~0.6\%. 
As~alternative background models for \jpsi\pip\pim candidates, 
a~first order polynomial function and a~product of 
an~exponential function with a~first order polynomial 
function are used. 
The~largest deviation of 
the~signal yield is found to be 1.5\%. 
The~two deviations are added in quadrature 
to obtain a 1.6\% systematic uncertainty due to imperfect
knowledge of the signal and background shapes.

Finally, the~finite size of the~simulation samples 
contributes an~uncertainty of~$0.9\%$ 
on the~ratio of total efficiencies.
The~total systematic uncertainty 
for the~ratio of branching fractions $\mathcal{R}$
is calculated as 
the~sum in quadrature of all the~values 
listed above and is found to be 6.0\%.

\begin{table}[tb]
\caption{Summary of systematic uncertainties
on 
the~ratio of branching 
fractions $\mathcal{R}$.
The~overall systematic uncertainty is calculated 
as a~sum 
in quadrature of all the sources.}
\label{tab:syst}
\begin{center}
\begin{tabular}{lc}
Source                           & Value [\%] 
\\[1mm]
\hline
\\[-2mm]
\Bp~kinematics 
& 0.1        \\
\Bu~decay model 
& 1.1        \\
Tracking efficiency correction   & 0.7        \\
Photon reconstruction correction & 3.6        \\
Kaon   identification            & 2.8        \\
Trigger efficiency               & 1.1        \\
Data-simulation agreement        & 3.0        \\
Fit model                        & 1.6        \\
Simulation sample size           & 0.9        
\\[1mm]
\hline
\\[-2mm]
Total & 6.0       
\end{tabular}\end{center} 
\end{table}

The~statistical significance 
for the~\mbox{$\decay{\Bu}{\jpsi\etapr\Kp}$}~decay
is recalculated using Wilks' theorem 
for each alternative fit model, 
and the~smallest values of 
$17$ and $10$~standard
deviations
for the~\etaprhog~and
\etapetapipi~cases, respectively, 
are taken as the~significance 
including 
the~systematic uncertainty.

\section{Results and summary}
\label{sec:sum}
The \betapk decays 
are observed for the first time
using 
proton\nobreakdash-proton collision data
collected by the~LHCb experiment  
at centre-of-mass energies of 
7, 8, and 13\tev,
corresponding 
to a~total integrated luminosity of $9\invfb$.
In~the~analysis, 
the~\etapr meson is reconstructed 
from the~two
$\rhoz\g$ and $\Peta\pip\pim$~final states.
The~signal significance 
exceeds 10 standard deviations
for both modes.
The~branching fraction of the~\betapk~decay 
is measured  for the~$\etaprhog$~sample 
through 
normalisation to 
the~known branching fraction of the~\bpsik~decay~\cite{PDG2022}.
The~ratio of branching fractions is found to be
\begin{equation*}
\dfrac{\BF(\betapk)}{\BF(\bpsik)} = 
\left(4.91\pm 0.47\pm0.29\pm0.07\right)\times10^{-2},
\end{equation*}
where the~first uncertainty is statistical, 
the~second  is systematic and the third  is related 
to 
the~uncertainties on 
the~branching fractions of 
the~intermediate resonances.
The~absolute branching fraction is determined using 
the~known branching fraction of \bpsik decays, 
\mbox{$\BF(\bpsik) = 
\left(6.24\pm0.20\right)\times10^{-4}$}~\cite{PDG2022},
and is found to be
\begin{equation*}
\BF(\betapk) = \left(3.06\pm0.29\pm0.18\pm0.04\right)\times10^{-5}\,,
\end{equation*}
where the~first uncertainty is statistical, 
the~second is systematic and 
the~third is due to external 
branching fractions uncertainties.
    The~measured branching fraction
    is consistent with the~upper limit
    previously set by the~\belle collaboration~\cite{Belle:2006aam}.
An~inspection of the~$\jpsi\etapr$~mass spectrum 
shows no significant contributions from 
the~decays via intermediate   charmonium or 
charmonium\nobreakdash-like resonances.


\section*{Acknowledgements}
%
%
\noindent We express our gratitude to our colleagues in the CERN
accelerator departments for the excellent performance of the LHC. We
thank the technical and administrative staff at the LHCb
institutes.
We acknowledge support from CERN and from the national agencies:
CAPES, CNPq, FAPERJ and FINEP (Brazil); 
MOST and NSFC (China); 
CNRS/IN2P3 (France); 
BMBF, DFG and MPG (Germany); 
INFN (Italy); 
NWO (Netherlands); 
MNiSW and NCN (Poland); 
MCID/IFA (Romania); 
MICINN (Spain); 
SNSF and SER (Switzerland); 
NASU (Ukraine); 
STFC (United Kingdom); 
DOE NP and NSF (USA).
We acknowledge the computing resources that are provided by CERN, IN2P3
(France), KIT and DESY (Germany), INFN (Italy), SURF (Netherlands),
PIC (Spain), GridPP (United Kingdom), 
CSCS (Switzerland), IFIN-HH (Romania), CBPF (Brazil),
Polish WLCG  (Poland) and NERSC (USA).
We are indebted to the communities behind the multiple open-source
software packages on which we depend.
Individual groups or members have received support from
ARC and ARDC (Australia);
Minciencias (Colombia);
AvH Foundation (Germany);
EPLANET, Marie Sk\l{}odowska-Curie Actions and ERC (European Union);
A*MIDEX, ANR, IPhU and Labex P2IO, and R\'{e}gion Auvergne-Rh\^{o}ne-Alpes (France);
Key Research Program of Frontier Sciences of CAS, CAS PIFI, CAS CCEPP, 
Fundamental Research Funds for the Central Universities, 
and Sci. \& Tech. Program of Guangzhou (China);
GVA, XuntaGal, GENCAT and Prog.~Atracci\'on Talento, CM (Spain);
SRC (Sweden);
the Leverhulme Trust, the Royal Society
 and UKRI (United Kingdom).




\clearpage 
\addcontentsline{toc}{section}{References}
\bibliographystyle{LHCb}
\bibliography{main,standard,LHCb-PAPER,LHCb-CONF,LHCb-DP,LHCb-TDR}

\newpage
\centerline
{\large\bf LHCb collaboration}
\begin
{flushleft}
\small
R.~Aaij$^{32}$\lhcborcid{0000-0003-0533-1952},
A.S.W.~Abdelmotteleb$^{51}$\lhcborcid{0000-0001-7905-0542},
C.~Abellan~Beteta$^{45}$,
F.~Abudin{\'e}n$^{51}$\lhcborcid{0000-0002-6737-3528},
T.~Ackernley$^{55}$\lhcborcid{0000-0002-5951-3498},
B.~Adeva$^{41}$\lhcborcid{0000-0001-9756-3712},
M.~Adinolfi$^{49}$\lhcborcid{0000-0002-1326-1264},
P.~Adlarson$^{77}$\lhcborcid{0000-0001-6280-3851},
H.~Afsharnia$^{9}$,
C.~Agapopoulou$^{13}$\lhcborcid{0000-0002-2368-0147},
C.A.~Aidala$^{78}$\lhcborcid{0000-0001-9540-4988}
% ...

{\footnotesize \it

% ...
$^{9}$Universit{\'e} Clermont Auvergne, CNRS/IN2P3, LPC, Clermont-Ferrand, France\\
% ...
$^{13}$LPNHE, Sorbonne Universit{\'e}, Paris Diderot Sorbonne Paris Cit{\'e}, CNRS/IN2P3, Paris, France\\
% ...
$^{32}$Nikhef National Institute for Subatomic Physics, Amsterdam, Netherlands\\
% ...
$^{41}$Instituto Galego de F{\'\i}sica de Altas Enerx{\'\i}as (IGFAE), Universidade de Santiago de Compostela, Santiago de Compostela, Spain\\
% ...
$^{45}$Physik-Institut, Universit{\"a}t Z{\"u}rich, Z{\"u}rich, Switzerland\\
% ...
$^{49}$H.H. Wills Physics Laboratory, University of Bristol, Bristol, United Kingdom\\
% ...
$^{51}$Department of Physics, University of Warwick, Coventry, United Kingdom\\
% ...
$^{54}$School of Physics and Astronomy, University of Glasgow, Glasgow, United Kingdom\\
$^{55}$Oliver Lodge Laboratory, University of Liverpool, Liverpool, United Kingdom\\
% ...
$^{63}$Syracuse University, Syracuse, NY, United States\\
% ...
$^{77}$Department of Physics and Astronomy, Uppsala University, Uppsala, Sweden, associated to $^{54}$\\
$^{78}$University of Michigan, Ann Arbor, MI, United States, associated to $^{63}$\\
\bigskip
% ...
}
\end{flushleft}

\end{document}